\documentclass[mathleft
]{an}
\usepackage{graphicx}
\usepackage{rotating}
\usepackage{lscape}
\usepackage{longtable}
\usepackage{times}
\begin{document}

\sloppy

\Pagespan{789}{}
\Yearpublication{2011}%
\Yearsubmission{2010}%
\Month{11}%
\Volume{999}%
\Issue{88}%

\title{Variations of $^{14}$C around AD 775 and AD 1795 -- due to solar activity}

\author{R. Neuh\"auser\inst{1} \thanks{Corresponding author: \email{rne@astro.uni-jena.de}}
\and D.L. Neuh\"auser\inst{2}
}

\titlerunning{14-C variation around AD 775}
\authorrunning{Neuh\"auser \& Neuh\"auser}

\institute{
Astrophysikalisches Institut und Universit\"ats-Sternwarte, FSU Jena,
Schillerg\"a\ss chen 2-3, 07745 Jena, Germany
\and
Schillbachstra\ss e 42, 07743 Jena, Germany
}

\received{23 Mar 2015}
\accepted{Augyst 2015}
\publonline{ }

\keywords{solar activity -- solar wind -- radiocarbon -- aurorae -- AD 775 -- AD 1795 -- BC 671 -- Dalton minimum}

\abstract{
The motivation for our study is the disputed cause 
for the strong variation of $^{14}$C around AD 775. 
Our method is to compare 
the $^{14}$C variation around AD 775 with other periods of strong variability.
Our results are: (a)
We see three periods, where $^{14}$C varied over 200 yr in a special way
showing a certain pattern of strong secular variation:
after a Grand Minimum with strongly increasing $^{14}$C, there is a series of
strong short-term drop(s), rise(s), and again drop(s) within 60 yr, ending up to 200 yr after
the start of the Grand Minimum. 
These three periods include the strong rises around BC 671, AD 775, and AD 1795.
(b) We show with several solar activity proxies (radioisotopes, sunspots, and aurorae)
for the AD 770s and 1790s that such intense rapid $^{14}$C increases
can be explained by strong rapid decreases in solar activity and, hence, wind,
so that the decrease in solar modulation potential leads to an increase in radioisotope production.
(c) The strong rises around AD 775 and 1795 are due to three effects, 
(i) very strong activity in the previous cycles (i.e. very low $^{14}$C level),
(ii) the declining phase of a very strong Schwabe cycle, 
and (iii) a phase of very weak activity after the strong $^{14}$C rise -- 
very short and/or weak cycle(s) like the suddenly starting Dalton minimum.
(d) Furthermore, we can show that the strong change at AD 1795
happened after a pair of two packages of four Schwabe cycles with certain hemispheric leadership 
(each package consists of two Gnevyshev-Ohl pairs, respectively two Hale-Babcock pairs). 
We show with several additional arguments that the rise around AD 775 was not that special.
We conclude that such
large, short-term rises in $^{14}$C (around BC 671, AD 775, and 1795) 
do not need to be explained by highly unlikely solar super-flares nor
other rare events, but by extra-solar cosmic rays modulated due to solar activity variations.
}

\maketitle

\section{Introduction: Strong rapid $^{14}$C variations}

Miyake et al. (2012, henceforth M12) found an increase in the $^{14}$C to $^{12}$C ratio
in two Japanese 
trees.
This increase was confirmed 
with trees on other continents 
(Usoskin et al. 2013, Jull et al. 2014, B\"untgen et al. 2014,
Wacker et al. 2014, G\"uttler et al. 2015).\footnote{This increase was
first noticed by McCormac et al. (2008) with 10-yr time resolution in Irish oak: {\em ... a rapid enrichment
of $^{14}$C between AD 765 and AD 775, which lies outside the $95\%$ probability envelope of Intcal94}
(while M12 found that their data and Intcal98 {\em are consistent with each other}).}
Even though $^{14}$C Intcal09 data have {\em 5-yr} time resolution for the last 11200 yr (Reimer et al. 2009), 
M12 searched for increases in {\em 10-yr} time 
steps. They
found three increases by more than 3 p.m. within 10 yr in the last three millenia,
namely BC 676-656, AD 760-785, and 1790-1820.
\footnote{As considered in M12, radiocarbon $^{14}$C does increase by more than
3 p.m. over 10 yr after adding up the increases in two subsequent 5-yr intervals around
both AD 775 and the 1790s; as far as the BC 676-656 period is concerned, the increase from
BC 676 to 656 is +6.7 p.m. over 20 yr, i.e. on average more than 3 p.m. over 10 yr. M12 gave here the period
{\em BC 675-655}, but the Intcal period is 2625 BP to 2605 BP (Before Present, i.e. before 1950) 
corresponding to the years $-675$ to $-655$
or BC 676 to 656, as there was no year $0$ in the BC/AD scale meaning {\em Before Christ} and {\em Anno Domini}.}

\subsection{Possible causes discussed so far}

For the AD 775 variation, M12 excluded supernovae as a possible cause due to the lack
of any historic observations and of any young nearby supernova remnants; they also excluded
solar super-flares, because they would not be frequent and hard enough to explain the
observed $^{14}$C to $^{10}$Be production ratio (M12).
Then, Usoskin \& Kovaltsov (2012), Melott \& Thomas (2012), Thomas et al. (2013), and Usoskin et al. (2013)
suggested that one (or several) solar super-flare(s) 
in AD 774 with only $\ge 24^{\circ}$ beam size (Melott \& Thomas 2012)
may have caused the 
event. They also estimated that 
4 to 6 times less $^{14}$C was
produced than given in M12 due to a different carbon model (Usoskin et al. 2013).
One would then need to postulate a solar super-flare which was only a few times larger 
than the so-called Carrington event (AD 1859),
but the Carrington event is not detected in radioisotopes (e.g., Neuh\"auser \& Hambaryan 2014).
It is clear that the Carrington event and the 1956 solar particle event, both undetected
in radioisotopes and cosmic rays, may have had different spectra than an event in 
AD 774/5. However, it
remains a fact that a solar flare was never detected in annual tree ring radiocarbon data --
Melott \& Thomas (2012), Usoskin et al. (2013), and Zhou et al. (2014) 
claim that AD 774/5 would be the first such case.

If the $^{10}$Be peak in the AD 780s (Horiuchi et al. 2008) and the $^{14}$C increase around AD 775
would be due to the same cause, and if 4-6 times less $^{14}$C was produced than estimated by M12
(Usoskin et al. 2013), then 45-68 times more $^{14}$C was produced than $^{10}$Be
(scaling from Hambaryan \& Neuh\"auser 2013).
According to calculations by Usoskin \& Kovaltsov (2012), $\sim 56$ times more $^{14}$C than $^{10}$Be
can be produced in solar flares, while according to Pavlov et al. (2014), only 16-17 times
more $^{14}$C than $^{10}$Be can be produced in solar flares.

Theoretically, the largest possible magnetic storm was estimated from equating the effective plasma pressure of the
solar magnetosheet and the magnetic pressure of the Earth dipole field to be $\sim 1.4$ times stronger than
the Carrington flare (Vasyliunas 2011); even when allowing a factor of $\sim 1.5$ more to account for
induction effects (Vasyliunas 2011), the largest flare would be only $\sim 2$ times larger than the Carrington event.
Also, from the largest observed solar magnetic field strength of 3500 G in the umbra and the
largest observed spot group with an area of 6000 MSH (millionth of a solar hemisphere),
Aulanier et al. (2013) have estimated the largest possible solar flare energy to be $\sim 6 \cdot 10^{33}$ erg,
i.e. much smaller than a super-flare; their scaling of flare energy and spot size would indicate
a single spot pair with $48^{\circ}$ extention for a flare energy of $10^{38}$ erg. Such a large
spot would have been observable since millenia at sunrise and sunset, but was never reported.
Other presumably solar-type stars showing super-flares may well be non-sun-like due to
faster rotation, much younger age, much larger surface, binarity, much higher magnetic fields, etc. 
(see e.g. Aulanier et al. 2013, Kitze et al. 2014, Wichmann et al. 2014).
See, e.g., Benz (2008) for a review on solar flares.
Cliver et al. (2014) also raised doubts 
on the solar super-flare hypothesis by comparison with the 1956 solar proton event.

Neuh\"auser \& Neuh\"auser (2015a) -- and also Stephenson (2015) in a smaller, less
complete study -- found that there were neither (strong) aurorae (in contrast to what
would have been expected for solar super-flares),
nor other extraordinary astrophysical events like supernovae, kilo-novae,
or gamma-ray-bursts in the mid 770s.
This result is based on a broad study of a well-documented period with sources
from Europe as well as Western and Eastern Asia. 

Hambaryan \& Neuh\"auser (2013) suggested a short gamma-ray-burst in AD 774 as cause,
as none of the observables including the $^{14}$C to $^{10}$Be production ratio is inconsistent with
such a burst (as confirmed by Pavlov et al. 2013)
-- even though a Galactic gamma-ray-burst may be unlikely given their 
highly uncertain cosmologic rate.
Kovaltsov et al. (2014) recently argued that the bi-modal
distribution in the cumulative occurence probability density function of annual fluence
(producing radioisotopes on Earth) could be due to solar activity (many events with low fluence)
and Galactic gamma-ray burst (few events with large fluence). However, the event rate of what
they interprete as Galactic gamma-ray bursts would range from one event in 100 to 9000 yr,
which is probably far too often (only one nearby Galactic gamma-ray burst beamed at us 
is expected in $3750 ^{+2250} _{-1442}$ kyr, see discussion in Hambaryan \& Neuh\"auser 2013). 
However, the cosmologic rate of gamma-ray burst is highly 
uncertain, see discussion in Hambaryan \& Neuh\"auser (2013). It can hardly be excluded
that one such event happened within the last millenia.

Liu et al. (2014) suggested a comet impact on Earth on AD 773 Jan 17 as cause for the 
radioisotope increases, allegedly seen in $^{14}$C in corals off the Chinese coast,
but this claim was falsified by several arguments.\footnote{Chapman et al. (2014)
have shown that the Chinese text incorrectly cited and translated by Liu et al. actually describes
a normal comet {\em with long tail} seen in China and Japan since AD 773 Jan 17 and 20, respectively --
the text does not mention an impact nor a bolide;
Usoskin \& Kovaltsov (2014) and Melott (2014) argued that comets cannot deliver such large amounts
of radioisotopes; Neuh\"auser \& Hambaryan (2014) could confirm that even the Tunguska event did not
lead to increased radioisotopes; the latter also argued that corals incorporate $^{14}$C a few
years after trees, so that an alleged $^{14}$C spike in coral in AD 773 Jan cannot correspond
to a $^{14}$C spike in trees in AD 775. The increase in those coral $^{14}$C data --
datable with $^{230}$Th to $\pm 14$ yr only -- 
may possibly be $\sim 2.5$ yr later than in $^{14}$C data from trees,
while in Liu et al. (2014), the data were shifted in time, so that the increase coincided
with a Chinese comet observation.}

\subsection{Was the AD 775 variation exceptional~?}

M12 wrote
that the {\em rapid increase of about 12 p.m. in the $^{14}$C content from AD 774 to 775 ...
is about 20 times larger than the change attributed to ordinary solar modulation}.
M12 may have considered that the mean $^{14}$C variation during the rise and decrease
of a Schwabe cycle is $\sim 3$ p.m. 
and that the mean half length of a Schwabe cycle is $\sim 5$ 
yr. If 
we then compare the typical Schwabe cycle modulation with 
an increase of $^{14}$C by $\sim 12$ p.m. in 1 yr (as claimed for AD 774/5), 
such a large fast increase would indeed be $\sim 20$ times larger than normal
($(5/1) \times (12/3) \simeq 20$).  

Considering the exact values for the time around AD 774/5 
and their error bars\footnote{We obtained the amplitude and its error for that time in the following way: 
we first applied a 5-yr filter to the $^{14}$C data, 
i.e. we calculated the 5-yr average for the (bi-)annual data in Miyake et al. (2013ab),
but using the averaged Miyake et al. (2012, 2013a) data for the time from
AD 750-820. Then, 
we subtracted the 5-yr averages from the real data to remove variations longer than the Schwabe 
cycle. The 
offsets between the real Miyake et al. data and the 5-yr averages are 
mainly due to the Schwabe cycle modulation (it also works similar with, e.g., 10-yr 
bins). In
each 5-yr bin, there is typically either a strong positive or a strong negative 
deviation due to solar activity minima and 
maxima. These
these strong (positive or negative) deviations can then be seen as Schwabe cycle amplitude in $^{14}$C.
We show the curve with the 5-yr averages in Fig. 4 compared to the Miyake et al. (2012, 2013ab) data,
which shows the positive and negative deviations 
and, hence, even though of some noise, the Schwabe cycle modulation in $^{14}$C
(shifted by a few yr due to the carbon 
cycle). A
8-13 yr band-pass filter ($\ge 95~\%$ of Schwabe cycles since
AD 1750 lie in the 8-13 yr range, Hathaway 2010, Schwabe cycle 4 being the only exception)
arrives at a similar result as our multi-proxy reconstruction of Schwabe cycles with $^{14}$C,
aurorae, and sunspots in Neuh\"auser \& Neuh\"auser (2015a).
For the time from AD 604 to 844 (excluding the strong variation from AD 774 to 784),
there are 22 Schwabe cycles:
the mean of the absolute values of the 22 strongest positive and the 22 strongest negative deviations 
(i.e. excluding the strong variations AD 774 to 784) is $2.4 \pm 1.2$ p.m. (AD 604-773 and 785-844),
which we see as mean Schwabe cycle half-amplitude in the Miyake et al. data.
The mean half-duration (rise and decrease) of a Schwabe cycle in those data is $5.22 \pm 0.77$ yr
obtained from the Miyake et al. (2013ab) $^{14}$C data from AD 604 to 844 without AD 774 to 784
by determining the time periods between local $^{14}$C maxima and minima; see 
also Neuh\"auser \& Neuh\"auser (2015a) for a more detailed discussion for the time AD 731 to 844.
The Schwabe cycle length and its $^{14}$C amplitude obtained by us for the time
around AD 774/5 have similar values and error bars in the last centuries (e.g., Stuiver 1998, Hathaway 2010).
}
the $11.9 \pm 2.3$ p.m. increase within
1 yr (AD 774/5) is only $2 \sigma$ deviant from the mean.

In addition, there are further intense increases within 1 yr time steps in Jull et al. (2014):
\begin{itemize}
\item at AD 762/763 an increase by $+7.3 \pm 4.9$ p.m. (Siberian tree),
\item at AD 792/793 an increase by $+7.9 \pm 4.9$ p.m. (Siberian tree),
\item at AD 768/769 an increase by $+9.1 \pm 5.0$ p.m. (Californian tree).
\end{itemize}
These increases have to be compared 
to AD 773/774 by $+6.7 \pm 4.9$ p.m. and AD 774/775 by $+9.2 \pm 6.0$ p.m. for the Siberian tree 
and, for the Californian tree, by $+13.9 \pm 4.9$ p.m. in AD 774/5 (all in Jull et al. 2014 data).
All these increases are consistent with each other within $1 \sigma$.
The advantage of the data by Jull et al. (2014) is the fact that they took data for
50 yr with 1 yr time-resolution, while, e.g., M12 had mostly 2-yr resolution. 
Furthermore, $^{14}$C around AD 774/5 in M12, Usoskin et al. (2013), 
and Jull et al. (2014) do increase for up to 4 yr.

\begin{figure*}[t]
\vspace{-0.15cm}
\begin{center}
{\includegraphics[width=10cm,angle=270]{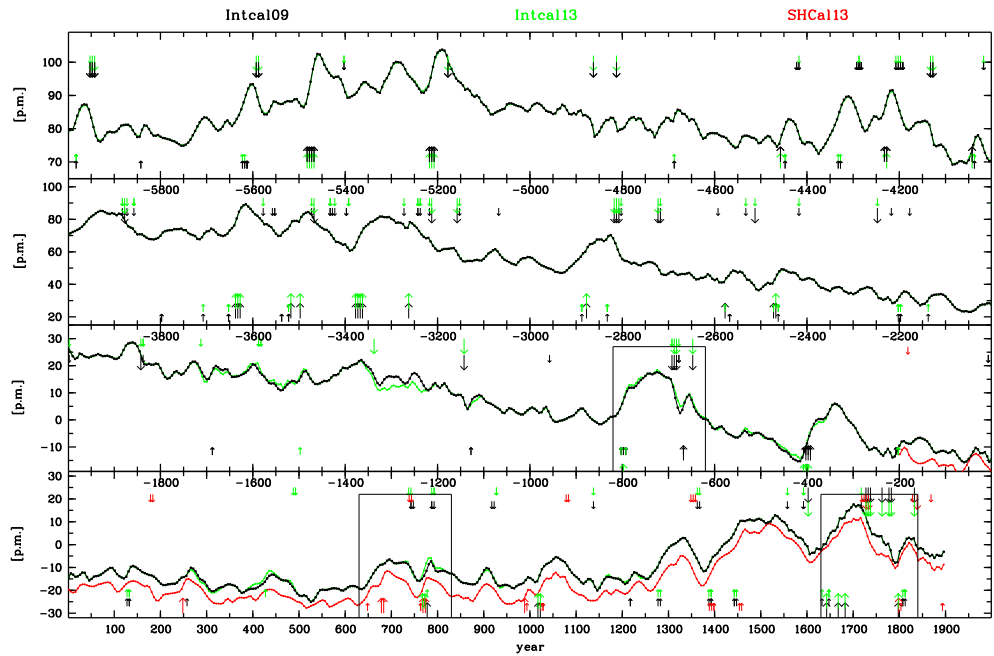}}
\caption{{\bf Radiocarbon $^{14}$C variations over the last $\sim 8000$ yr}
before the Suess effect in AD 1900 from Intcal09 (black, Reimer et al. 2009),
Intcal13 (green, Reimer et al. 2013), and SHCal13 (red, Hogg et al. 2013b), in p.m. in 5-yr time steps;
SHCal13 is shown only after BC 200, since it is essentially like Intcal13 before that time,
but only shifted by the interhemispheric offset (Hogg et al. 2013b).
The long down-facing arrows indicate strong rapid decreases by at least 2.5 p.m. in 5-yr steps in the respective
data set (indicated by its color), the long up-facing arrows indicate strong rapid increases by at least 
2.5 p.m. in 5-yr steps in the respective data set (indicated by its color).
The short arrows indicate further (weaker) decreases (and increases) by 2 to 2.5 p.m. in 5 yr, respectively.
Some arrows are plotted (partly) on top of others; also no arrows for SHCal13 before BC 200.
The arrows are plotted into the middle of the 5-yr bin of the incorporation epoch.
The boxes indidate the three special periods of $\le 200$ yr each studied in this paper,
namely periods with intense variations: first a Grand Minimum, then a sequence of 
strong rapid decreases, increases, and again decreases within 60 yr around the
AD 1795, AD 775, and BC 671, the only three such periods between BC 1000 and AD 1900.}
\end{center}
\end{figure*}

\subsection{Our approach}

If the variation at AD 774/5 was not that exceptional,
and -- even though of the carbon cycle --
if the underlying effect may have lasted a few years, 
the variation at AD 774/5 may have been more similar 
to other instances.
One should consider effects other than suggested so far,
namely lasting up to a few years.

\begin{table*}
\begin{center}
\caption{$^{14}$C {\em in}creases within one Intcal09 (Reimer et al. 2009), 
Intcal13 (Reimer et al. 2013), or SHCal13 (Hogg et al. 2013b) time step (5 yr) from 1000 BC to AD 1900,
which are at least $+2.5$ p.m. in 5 yr in at least one of those series, sorted here by $^{14}$C rises.
For those 5-yr time-steps, for which a significant rise was found only in one or two of the data sets,
we give the (smaller) rise(s) in the other set(s) in brackets
(a rounded value of 2.5 p.m. given in brackets is actually slightly below 2.5 p.m. before rounding).
The $^{14}$C increase from AD 1795-1800 is the largest one both in S/N and
in total $^{14}$C rise in p.m. in all three series.}
\begin{tabular}{lccccccl} \hline
Time           & \multicolumn{2}{c}{Intcal09} & \multicolumn{2}{c}{Intcal13} & \multicolumn{2}{c}{SHCal13} & comment \\
step           & $^{14}$C in 5 yr & S/N  & $^{14}$C in 5 yr & S/N  & $^{14}$C in 5 yr & S/N & \\ \hline 
AD 1795 - 1800 & $+4.8 \pm 1.4$    & 3.4  & $+4.6 \pm 1.3$    & 3.5 & $+3.2 \pm 1.6$    & 2.0 & 1790s \\
AD 1665 - 1670 & $+3.8 \pm 1.4$    & 2.7  & $+3.7 \pm 1.3$    & 2.9 & ($+1.9 \pm 2.1$)  & 0.9 & Maunder \\
AD 765 - 770   & ($+1.5 \pm 2.3$)  & 0.7  & $+3.2 \pm 2.0$    & 1.6 & $+3.2 \pm 2.5$    & 1.3 & AD 770s (*) \\
AD 770 - 775   & ($+2.1 \pm 2.3$)  & 0.9  & $+3.2 \pm 2.1$    & 1.5 & $+2.5 \pm 2.4$    & 1.0 & AD 770s \\
BC 406 - 401   & $+2.9 \pm 2.2$    & 1.3  & $+3.0 \pm 2.1$    & 1.4 & (****) & 0.7 & \\
BC 401 - 396   & $+2.8 \pm 2.3$    & 1.2  & $+2.9 \pm 2.1$    & 1.4 & (****) & 0.7 & \\
BC 396 - 391   & $+2.5 \pm 2.4$    & 1.0  & $+2.9 \pm 2.1$    & 1.4 & (****) & 0.7 & \\ 
AD 775 - 780   & $+2.9 \pm 2.3$    & 1.3  & ($+2.1 \pm 2.1$)  & 1.0 & ($+0.2 \pm 2.3$)  & 0.1 & AD 770s \\
AD 1680 - 1685 & $+2.8 \pm 1.4$    & 2.0  & $+2.7 \pm 1.3$    & 2.1 & ($+0.3 \pm 1.6$)  & 0.1 & Maunder \\
BC 801 - 796   & ($+2.0 \pm 2.2$)  & 0.9  & $+2.7 \pm 2.1$    & 1.3 & (****) & 0.6 & BC Gr. Min.\\
AD 1020 - 1025 & $+2.7 \pm 2.2$    & 1.2  & $+2.5 \pm 2.0$    & 1.3 & ($+2.2 \pm 2.3$)  & 1.0 & Oort \\
BC 671 - 666   & $+2.7 \pm 2.3$    & 1.2  & $+1.8 \pm 1.8$    & 1.0 & (****) & 0.4 & 7th c. BC \\
AD 1640 - 1645 & $+2.6 \pm 1.4$    & 1.9  & $+2.6 \pm 1.3$    & 2.0 & ($+0.8 \pm 2.2$)  & 0.4 & Maunder \\
AD 1015 - 1020 & $+2.6 \pm 2.2$    & 1.2  & ($+2.5 \pm 2.0$)  & 1.3 & ($+1.7 \pm 2.2$)  & 0.8 & Oort \\
AD 675 - 680   & ($+0.2 \pm 2.3$)  & 0.1  & ($+0.5 \pm 2.3$)  & 0.2 & $+2.6 \pm 2.4$    & 1.1 & Dark Age (**) \\
AD 985 - 990   & ($+1.3 \pm 2.0$)  & 0.7  & ($+1.3 \pm 2.0$)  & 0.7 & $+2.5 \pm 2.0$    & 1.3 & (***) \\
BC 411 - 406   & ($+2.3 \pm 2.3$)  & 1.0  & $+2.5 \pm 2.1$    & 1.2 & (****) & 0.5 & \\ 
AD 245 - 250   & ($+0.2 \pm 2.2$)  & 0.1  & ($+0.6 \pm 2.1$)  & 0.3 & $+2.5 \pm 2.3$    & 1.1 & \\
AD 680 - 685   & ($+0.5 \pm 2.4$)  & 0.2  & ($+0.5 \pm 2.3$)  & 0.2 & $+2.5 \pm 2.4$    & 1.0 & Dark Age (**) \\ \hline
\end{tabular}
\end{center}

Remarks: 
See also footnotes 5 \& 6.
(*) An apparently strong rise from AD 765-770 (in Intcal13 and SHCal13) is not supported
by the data with 1-yr time-resolution, see below, Fig. 9; that this rise appears in Intcal13, but not
in Intcal09, is probably due to the inclusion of the data by McCormac et al. (2008) with 10-yr
time-resolution and interpolation in Intcal; that this rise appears in SHCal13 might be due
to a small (few yr) calendar age offset between Intcal north and SHCal south, see Sect. 4.2.
(**) There is an additional increase by $+2.4$ p.m. in SHCal13 from AD 645 to 650 (Fig. 1),
plus increases by $+1.7$ to $+1.8$ p.m. in Intcal09 and Intcal13 from
AD 645 to 650 to 655 to 660.
(***) This $^{14}$C rise 
is close in time to the increase found in a Japanese tree 
from $-20.7 \pm 1.6$ p.m. in AD 993 to $-11.5 \pm 2.0$ p.m. in AD 994, i.e. by $+9.2 \pm 2.6$ p.m. (Miyake et al. 2013a); 
date corrected in the Corrigendum 2013 Nov 7 (Miyake et al.); this rapid $^{14}$C rise AD 993/4 was previously seen
in Menjo et al. (2005) from AD 994-996 by $+6.3 \pm 2.8$ p.m. 
(****) Before BC 200, SHCal13 is essentially like Intcal13, but shifted only by the interhemispheric offset (Hogg et al. 2013b),
so that those data do not provide additional or independant information.
\end{table*}

\begin{table*}
\begin{center}
\caption{$^{14}$C {\em de}creases within one Intcal09, Intcal13, and SHCal13 time step (5 yr) 
from 1000 BC to AD 1900, which are at least $-2.5$ p.m. in 5 yr, sorted by $^{14}$C amplitude.
For those 5-yr time-steps, for which a significant decrease was found only in one or two of the data sets,
we give the (smaller) decrease in the other set(s) in brackets
(a rounded value of -2.5 p.m. given in brackets is actually slightly above -2.5 p.m. before rounding).
Strong $^{14}$C decreases in Intcal09 are found mostly before the intense increase in the 7th century BC
and in the 18th century AD before the strong rapid increase in the 1790s.
There are five more such strong decreases from 1905 to 1945 just due to the Suess effect,
i.e. not listed here.}
\begin{tabular}{lccccccl} \hline
Time           & \multicolumn{2}{c}{Intcal09} & \multicolumn{2}{c}{Intcal13} & \multicolumn{2}{c}{SHCal13} & comment \\
step           & $^{14}$C in 5 yr & S/N  & $^{14}$C in 5 yr & S/N  & $^{14}$C in 5 yr & S/N & \\ \hline
AD 1780 - 1785 & $-4.4 \pm 1.3$   & 3.4  & $-4.3 \pm 1.3$   & 3.3  & ($-1.9 \pm 1.6$) & 1.2  & 18th century \\
BC 691 - 686   & $-4.1 \pm 2.4$   & 1.7  & $-3.0 \pm 1.8$   & 1.7  & (**)  & 0.7  & 7th c. BC \\
AD 1725 - 1730 & $-3.7 \pm 1.3$   & 2.8  & $-3.6 \pm 1.3$   & 2.8  & $-3.3 \pm 2.1$   & 1.6  & 18th century \\
BC 686 - 681   & $-3.6 \pm 2.4$   & 1.5  & $-3.6 \pm 2.4$   & 1.5  & (**)  & 0.6  & 7th c. BC \\
AD 1600 - 1605 & $-3.5 \pm 1.3$   & 2.7  & $-2.7 \pm 1.8$   & 1.5  & ($-0.9 \pm 2.4$) & 0.4  & \\
AD 1835 - 1840 & ($-1.7 \pm 1.4$) & 1.2  & ($-1.7 \pm 1.4$) & 1.2  & $-3.4 \pm 1.6$   & 2.1  & end Dalton \\
AD 1830 - 1835 & $-3.0 \pm 1.4$   & 2.1  & $-2.9 \pm 1.3$   & 1.3  & ($-1.4 \pm 1.8$) & 0.8  & end Dalton \\
AD 1775 - 1780 & $-2.9 \pm 1.3$   & 2.2  & $-2.8 \pm 1.3$   & 2.2  & ($-1.7 \pm 1.6$) & 1.1  & 18th century \\
BC 696 - 691   & $-2.8 \pm 2.4$   & 1.2  & ($-2.0 \pm 2.0$) & 1.0  & (**) & 0.6  & 7th c. BC \\
AD 1760 - 1765 & $-2.7 \pm 1.3$   & 2.1  & $-2.7 \pm 1.3$   & 2.1  & ($-0.8 \pm 2.1$) & 0.4  & 18th century \\
BC 651 - 646   & $-2.6 \pm 2.1$   & 1.2  & $-2.5 \pm 1.9$   & 1.3  & (**)  & 0.6  & 7th c. BC \\
AD 1730 - 1735 & $-2.6 \pm 1.4$   & 1.9  & $-2.5 \pm 1.4$   & 1.8  & $-2.8 \pm 2.1$   & 1.3  & 18th century \\
AD 1735 - 1740 & $-2.5 \pm 1.4$   & 1.8  & $-2.5 \pm 1.3$   & 1.9  & ($-1.9 \pm 2.1$) & 0.9  & 18th century \\ \hline
\multicolumn{8}{c}{Additional, but weaker strong rapid decreases in 8th century AD (*):} \\ \hline
AD 735 - 740   & ($-1.4 \pm 2.3$)  & 0.6  & ($-2.4 \pm 2.1$) & 1.1 & ($-2.1 \pm 2.4$)  & 0.9  & 8th century \\
AD 740 - 745   & ($-2.0 \pm 2.3$)  & 0.9  & ($-2.4 \pm 2.0$) & 1.2 & ($-2.0 \pm 2.3$)  & 0.9  & 8th century \\
AD 745 - 750   & ($-2.2 \pm 2.3$)  & 1.0  & ($-1.9 \pm 1.9$) & 1.1 & ($-1.4 \pm 2.3$)  & 0.6  & 8th century \\
AD 785 - 790   & ($-2.0 \pm 2.3$)  & 1.0  & ($-2.4 \pm 2.0$) & 1.2 & ($-0.4 \pm 2.3$)  & 0.2  & 8th century \\
AD 790 - 795   & ($-2.2 \pm 2.2$)  & 1.0  & ($-2.0 \pm 2.0$) & 1.0 & ($-0.4 \pm 2.3$)  & 0.2  & 8th century \\ \hline
\end{tabular}
\end{center}

Remarks: 
See also footnotes 5 \& 6. 
(*) When releasing the selection to include decreases by $-2.0$ to $-2.5$ p.m. in 5 yr,
then one gets 15 more {\em events} before AD 1900 including the strong decreases 
at AD 735-740, 740-745, 745-750, 785-790, and 790-795, which are listed in the bottom part.
(**) Before BC 200, SHCal13 is essentially like Intcal13, but shifted only by the interhemispheric offset (Hogg et al. 2013b),
so that those data do not provide additional or independant information.
\end{table*}

We will study here whether more or less normal solar activity variation can cause
such a strong $^{14}$C increase. If the AD 770s rise is not that special,
then let us search for similar $^{14}$C variations in the last few millenia.

We first search for $^{14}$C variations from BC 1000 to AD 1900,
which may have been similar as around AD 775 (Sect. 2).
Then, we explain 
the solar activity proxies used in this study 
(Sect. 3). Next,
we present the variation of those proxies
around BC 671 as well as in the AD 770s and the 1790s 
in Sect. 4.
We discuss our results quantitatively in terms of solar modulation potential
and $^{14}$C production rate 
in Sect. 5.
Finally, we discuss the timing of the increases with respect to the state of solar activity
and the solar magnetic field
in Sect. 6 and conclude with a summary of the main results (Sect. 7).

\section{The largest increases and decreases in Intcal data (BC 1000 -- AD 1900)}

While Intcal09 had a 5-yr time resolution back to 11,200 BP (Reimer et al. 2009),
Intcal13 (north, Reimer et al. 2013) and SHCal13 (south, Hogg et al. 2013b) 
both have this time resolution back to 13,900 BP;
some of the revisions from Intcal09 to Intcal13 concern the last three millenia
(including even the AD 770s).
We have searched for the largest increases and decreases 
in 5-yr time steps 
not only in Intcal09, but 
also in Intcal13 and SHCal13 (arrows in Fig. 1).
We list all intense rapid increases by at least $+2.5$ p.m in a 5-yr-step
and decreases by at least $-2.5$ p.m in a 5-yr-step
from 1000 BC to AD 1900 in Tables 1 and 2, respectively.\footnote{We also searched for increases (and decreases)
by $+2.5$ to $+2.0$ p.m. (and $-2.0$ to $-2.5$ p.m., respectively) in a 5-yr-step, in order to be able
to compare with Usoskin \& Kovaltsov (2012), who searched for increases in Intcal09 by 
more than 2 p.m. in up to 10 yr with a decrease (with a similar amplitude in p.m.) afterwards showing a curve
with FWHM of 15-30 yr (to resemble the curve expected after a short flare due to the carbon cycle);
for the time since 1000 BC, they list two cases (8th and 3rd century AD), 
e.g. for AD 255-260 there is an increase by $+2.1$ p.m. in Intcal09, 
with the largest decrease in the next few decades being $-1.8$ p.m. from AD 275-280
(i.e. neither in our Table 2 nor our Fig. 1);
for the period around BC 676-656, one of the three large variations mentioned in M12,
Usoskin \& Kovaltsov (2012) do not list a large increase (followed by a large decrease),
but there are rises by $\ge +2.0$ p.m. within 5 and 10 yr from BC 676-656 in Intcal09
(e.g. $+2.0$ p.m. from BC 666-661, $+2.7$ p.m. from BC 671-666, or $+3.9$ p.m. from BC 676-666) 
followed by a decrease by $-2.6$ p.m. from BC 651-646.}

The largest increase is found at AD 1795-1800 with $+4.8 \pm 1.4$ p.m.
The variations at
AD 775-780 and BC 671-666 were also among
the nine largest increases (in Intcal09) within 5-yr time steps, see Table 1.\footnote{Note that
radiocarbon is incorprated into northern hemisphere tree rings
in the northern spring and summer, i.e. around the middle of a calendar year
(e.g. for year $x$ or $775$ at epoch $x.5$ or $775.5$),
while for the southern hemisphere,
the incorporation takes place in the southern spring and summer,
i.e. around the turn of the year. Conventionally, a northern hemisphere tree ring is labeled $x$ or $775$,
with the incorporation being at $x.5$ or $775.5$;
a southern hemisphere tree ring is labeled $x$ or $775$,
even though the incorporation was at $x+1.0$ or $776.0$, the Schulman (1956) southern hemisphere dating convention.
For the data in Tables 1 and 2 and whenever we give the time range of
a 5-yr-step from those tables, we list the years as in the Intcal and SHCal13 publications
(but in the AD/BC scale instead of the BP scale), unless otherwise specified,
i.e. we list the tree ring years as labeled in the publications, but not the exact epoch of carbon incorporation.
In the figures, though, we consider the difference between label and incorporation year
and always specify whether we plot either
into the incorporation epoch, e.g. Fig. 1, or shift the data by 2.5 yr
from the incorporation to the past due to the carbon cycle, see below.}

Three $^{14}$C variations are special in the following sense: \\
First, there is a Grand Minimum  
with one or more strong rapid increases (by $\ge +2.0$ p.m. in 5 yr).
Then, afterwards the scenario is like that: 
a fall followed by a rise followed by a fall.
More precisely:
one or more strong rapid decreases (by $\le -2.0$ p.m. in 5 yr) 
followed by at least one intense rapid rise ($\ge +2.0$ p.m. in 5 yr) 
and then again followed by one or more strong rapid decreases 
(by $\le -2.0$ p.m. in 5 yr), all within 60 yr.
From the start of the preceeding Grand Minimum to the last 
strong rapid decrease, there are only up to 200 yr.
Those three cases are the only ones between 1000 BC and AD 1900, they are indicated in Fig. 1
(there are a few somewhat similar cases in the 5000 yr before 1000 BC). 
These three cases show intense variability in solar activity over a certain time-scale (within 200 yr)
and therefore merit further investigations.
The strongest decrease (AD 1780-1785) and the strongest increase (AD 1795-1800) 
are even within 20 yr. The Earth magnetic field is not known nor believed to varry that fast,
so that those $^{14}$C variations are due to solar activity variations.
At least seven of the 19 strong increases in $^{14}$C listed in Table 1 happened during a
(known, named) Grand Minimum -- it is a consensus that solar activity drops strongly during a
Grand Minimum, and that there were less and weaker aurorae and flares during Grand
Minima than during normal activity.

Note that some aspects of these three strong de Vries variations were studied before:
Damon et al. (1978) and Damon (1988) for the AD 1790s, McCormac et al. (2008) and M12 for the AD 770s,
and Takahashi et al. (2007) for BC 671.
M12 considered the three increases as the largest within 10 yr time steps (see footnote 2).

\section{Solar activity proxies}

Here, we present the solar activity proxies use in this paper.

\subsection{Sunspots}

We use telescopic and naked-eye sunspots, aurorae, and also radioisotopes
to reconstruct solar activity at the times of intense rapid $^{14}$C variations
around AD 1795 and 775; for the time around BC 671, only radioisotopes are available. 

We use sunspot group numbers from Hoyt \& Schatten (1998) and naked-eye sunspot reports
from Vaquero et al. (2002) and references therein as well as J. Vaquero (priv. comm.).

\subsection{Aurorae}

For aurorae after AD 1700, we use the aurora catalogs 
from Fritz (1873), Tromholt (1902), Legrand \& Simon (1987), and Krivsky \& Pejml (1988). 

For the time from AD 670 to 845, we have compiled aurorae from the following references:
Jeremiah (1870), Fritz (1873), 
Matsushita (1956), Link (1962), Schove (1964, 1984), Newton (1972),
Keimatsu (1970, 1973, 1974), Keimatsu \& Fukushima (1976), Dall'Olmo (1979), 
Yau et al. (1995), Bone (1996), Hetherington (1996), McCarthy \& Breen (1997), 
Silverman (1998, with online catalog\footnote{This catalog lists all sources as
individual entries, so that many aurorae events are listed several times,
see Neuh\"auser \& Neuh\"auser (2015a) for more details.}, Xu et al. (2000),
Rada \& al-Najeh (1997), Basurah (2005, 2006), and Usoskin et al. (2013).
From Keimatsu (1970, 1973, 1974), Keimatsu \& Fukushima (1976), and Matsushita (1956),
we include all those with reliability 1-3.
Like in Xu et al. (2000), we do not include those presumable aurora events from Dai \& Chen (1980),
which either do not mention {\em night} or which mention a certain strange new star,
while Silverman (1998) does include such events. \\
For the time from AD 731 to 825, we have revised all reports thoroughly;
in Neuh\"auser \& Neuh\"auser (2015a), we explain the selection, identification,
and classification criteria, present the texts and individual discussions,
and we also list events previously misdated and/or misinterpreted as aurorae.
For the periods from AD 670 to 730 and AD 826 to 845, 
we have only excluded obvious non-auroral events.
For the period from AD 550 to 845, we will publish a compilation of most original texts 
and English translations for all aurora observations soon.
 
Recently, Stephenson (2015) published a compilation of possible aurorae from AD 767 to 779
with eight events from Eastern Asia (two are immediatelly excluded as aurorae 
by Stephenson 2015) and two from Europe.
The remaining eight events are also listed in our catalog from AD 731-825 (Neuh\"auser \& Neuh\"auser 2015a),
three as likely true aurorae and five as other events like halos.
In addition to those three likely true aurorae,
we (Neuh\"auser \& Neuh\"auser 2015a) list three more likely true aurorae in AD 772 and 773.
The two events possibly to be interpreted as aurorae according to Stephenson (2015)
near AD 775, namely 
{\em above the} (almost full) {\em moon ... more then ten streaks of white vapour} 
(from east to south) in AD 776 Jan in China 
and a {\em red cross ... in the heavens after sunset} in AD 776 in England
(often misdated to AD 773 or 774) are both more likely halo displays, see detailed discussion in
Chapman et al. (2015) and Neuh\"auser \& Neuh\"auser (2014, 2015ab).

\subsection{Radiocarbon}

For $^{14}$C, we use Intcal09 (Reimer et al. 2009), Intcal13 (Reimer et al. 2013),
and SHCal13 (Hogg et al. 2013b), all with 5-yr time resolution;
then also data with 1- to 5-yr time resolution from Stuiver et al. (1998), Miyake et al. (2012, 2013ab),
McCormac et al. (2008), Usoskin et al. (2013), Jull et al. (2014), B\"untgen et al. (2014),
Wacker et al. (2014)\footnote{Wood from a Swiss church which was considered to be cut in the AD 780s,
so that the strong $^{14}$C rise could constrain the age even further -- but these data are not
dated in an absolute sense as the others.}, and Takahashi et al. (2007)
(we did not use the Kocharov et al. (1995) et al. data, because they show unsual large
amplitudes and were not available with error bars); 
for the southern hemisphere, we also use data 
around AD 770s and 1790s from Stuiver \& Braziunas (1998),
Zimmerman et al. (2010), Hogg et al. (2011, 2013a), and G\"uttler et al. (2015).
For data from B\"untgen et al. (2014), Wacker et al. (2014), Stuiver \& Braziunas (1998),
Zimmerman et al. (2010), and Hogg et al. (2011, 2013a), we converted the published radiocarbon
content given as $^{14}$C age $^{14}$C in p.m. following Stuiver \& Polach (1977).

Solar activity modulates incoming cosmic rays, but with some delays 
(reviews in Hathaway 2010 and Usoskin 2013);
Hathaway (2010) wrote: {\em The reduction in cosmic ray flux tends to lag 
behind solar activity by 6- to 12-months (Forbush, 1954).}
When considering periods of increased radioisotope variations, we take into account differences
in atmospheric residence times and for incorporation of $^{14}$C into tree rings 
-- the carbon cycle -- and $^{10}$Be into ice layers:
while $^{14}$C has an atmospheric residence time of $\sim 20$-25 yr with most
$^{14}$C incorporated in tree rings within the first $\sim 2-3$ years (Sigman \& Boyle 2000),
$^{10}$Be is incorporated mostly within the first year (Heikkil\"a et al. 2008).
About 2-3 yr after the $^{14}$C production, there is the largest yearly amount of incorporation 
into tree rings, and there is also a significant incorporation in the very first year.
Therefore, we shifted the $^{14}$C data by 2.5 yr in most figures (unless otherwise specified),
as recommended by Houtermans et al. (1973), Braziunas (1990), and Stuiver (1994);
e.g., from AD 774.5, when the $^{14}$C was incorporated into the tree ring, to AD 772.0 as plotted;
for southern hemisphere data, the incorporation takes place in the southern spring and summer around 
the turn of the year, i.e. at $x.0$ (for tree ring of year x) 
or e.g. 774.0, so that we would shift by 2.5 yr to $x-2$ or 771.5;
when we mention years in the text, we specify whether we mean shifted or unshifted years,
whenever neccessary. 
After having shifted the $^{14}$C by those 2.5 yr, the Schwabe cycle activity maxima 
correspond well with $^{14}$C minima and viceversa.
We will see in Sect. 4.1 and Fig. 2, how fast the $^{14}$C reacted (within few years), 
so that our procedure is justified. 
Hence, $^{14}$C data shifted back by 2.5 yr are a good approximation
of the $^{14}$C production rate, which we do not use to avoid further assumptions.

\subsection{$^{10}$Be data}

For $^{10}$Be, we use GRIP (Vonmoos et al. 2006), Dye 3 (Beer et al. 1990),
NGRIP (Berggren et al. 2009), and Dome Fuji data (Horiuchi et al. 2008).

\section{Solar activity variations around BC 671, AD 775, and 1795}

We can now study the solar activity variation for the three 
periods under interest -- using all known proxies. 

Since there are neither sunspot nor aurora data available for the time around BC 671,
we concentrate our study on AD 770s and 1790s.
We start the discussion with the most recent one,
because most data are available here.

\subsection{Solar activity variations around AD 1795}

After almost thousand years without strong Grand Minima (Fig. 1),
there were several Grand Minima since about AD 600
(Dark Age, Oort, Wolf, Sp\"orer, and Maunder).
Since the end of the Maunder Grand Minimum, solar activity was 
relatively intense (Grand Maximum) -- except a few Schwabe
cycles like, e.g., in the short Dalton Minimum.
From about AD 600 to 1850, there were quite a number of strong
rapid $^{14}$C increases and decreases (Tables 1 and 2, Fig. 1).

\subsubsection{Increasing activity until around AD 1795}

The $^{14}$C evolution before the AD 1790s was as follows:
after the end of the Maunder Grand Minimum around AD 1712/1715,
solar activity increased over eight Schwabe cycles --
as it is seen clearly in sunspots, aurorae, and $^{14}$C (Figs. 2 \& 3). 

\begin{figure*}
\begin{center}
{\includegraphics[angle=270,width=15cm]{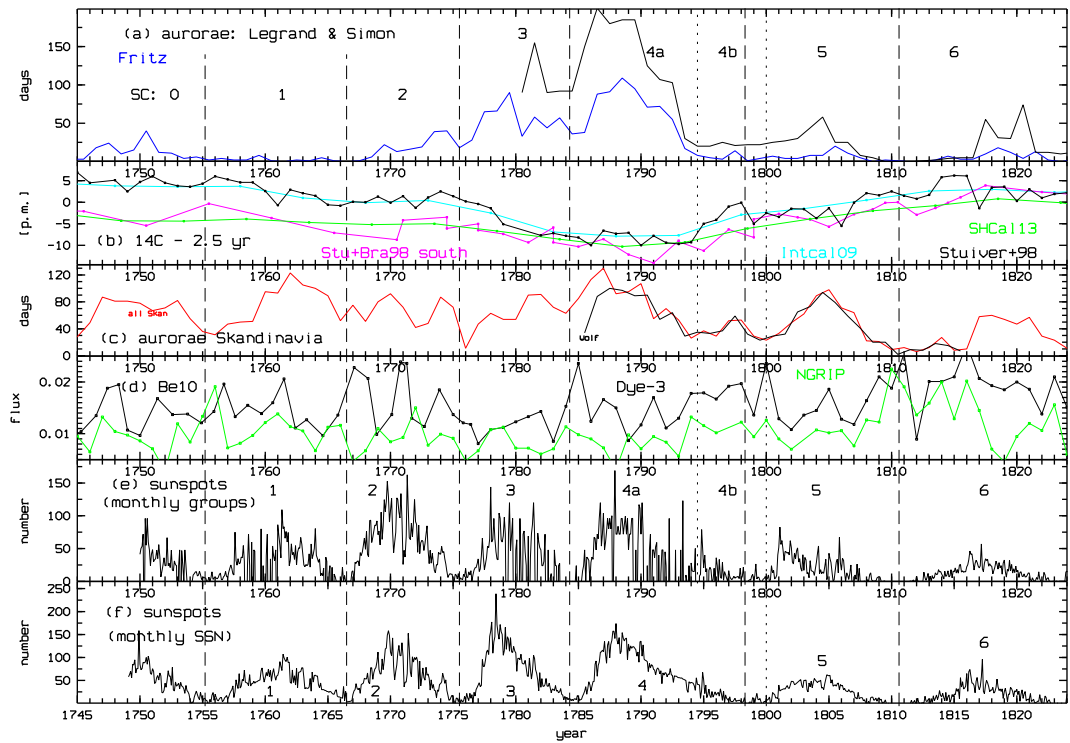}}
\caption{{\bf Solar activity proxies from 1745 to 1824:}
(a) Aurorae (days with aurora activity per year) from Legrand \& Simon (1987) 
(upper curve since 1780) and 
then those from Fritz (1873) as lower curve (blue);
there is an extra weak peak around AD 1796/7; all those aurorae are plotted in the middle 
of the respective year (e.g. at epoch $x.5$ for year $x$).
(b) $^{14}$C (in p.m.) from Stuiver et al. (1998) in black,
Intcal09 (Reimer et al. 2009) in light blue, 
both for the northern hemisphere, then SHCal13 (Hogg et al. 2013b) in green
and data from Stuiver \& Braziunas (1998) in pink for the southern hemisphere,
all shifted 2.5 yr (to the past) in this plot
(carbon cycle) to facilitate better comparison with the other proxies.
We can clearly see that aurorae (solar wind) and $^{14}$C are inverse proportional.
(c) Aurorae (days with aurora activity per year) from Tromholt (1902) 
all Skandinavia (red),
at a geographic latitude range of $55-68.5^{\circ}$N (his zones II to V),
which is always south of the (typical) polar auroral oval given the location of the geomagnetic pole at that time
(reconstructed to be located in AD 1800 at $78.85^{\circ}$N and $60.91^{\circ}$E by Korte \& Mandea 2008);
since Tromholt lists yearly aurora sums from July (in year $x-1$) to June of the next year ($x$),
we plot them at the turn of the calendar year at epoch $x.0$;
also shown are aurorae from Rubenson (as listed in Wolf 1880) in black 
(plotted in the middle of the respective year at epoch $x.5$);
the Tromholt and Rubenson aurorae show peaks for two Schwabe cycles: 
one broad maximum from 1787 to 1790 and one more peak 1797/8.
(d) $^{10}$Be flux (in 0.01 atoms cm$^{-2}$ s$^{-1}$) with 1-yr time resolution 
from Dye 3 (Beer et al. 1990) in black and NGRIP (Berggren et al. 2009) in green.
(e) Monthly sunspot group numbers (Hoyt \& Schatten 1998).
(f) Monthly sunspot numbers (SSN, Hathaway 2010) with the very long Schwabe cycle 4
(with few observations).
We also indicate the Schwabe cycle numbers. 
Borders between cycles are plotted as vertical dashed lines at the
sunspot minima from Hoyt \& Schatten (1998), 
but as dotted lines for the minima at the end of cycles 4a and 4b.
}
\end{center}
\end{figure*}

The eight Schwabe cycles no. -3 to 4a from 1712.0 until 1793.1
lasted 81.1 yr with a mean length of $9.1 \pm 1.6$ yr only
(or $10.8 \pm 1.3$ yr when leaving the long cycle no. 4 
undivided).\footnote{The strong, but long Schwabe cycle numbered no. 4
was suggested to be split into two shorter cycles (we call them 4a and 4b):
the extra Schwabe cycle was suggested by Loomis (1870) based on
aurorae and by Usoskin et al. (2001a) based on
new sunspot group numbers by Hoyt \& Schatten (1998),
it is not inconsistent with the newly reconstructed butterfly diagram
(Usoskin et al. 2009) using old sunspot observations found by Arlt (2009).
See Fig. 2 for the solar activity proxies around that time.
According to Usoskin et al. (2001a), the extra minimum between the two cycles inside
cycle no. 4 was in AD 1793.1, and the following next minimum was
either 1799.8 (Usoskin et al. 2001a) or 1798.3 (Hoyt \& Schatten 1998, Hathaway 2010),
resulting in a cycle length of only 5.2 to 6.7 yr for 4b.
According to tables 1 and 3 in Hathaway (2010), cycle no. 4
would have had a length of 17 to 18 yr (from maximum to maximum)
and a length of 14 yr from minimum to minimum.
The mean SC length from cycles 1 to 22 is $131 \pm 14$ month = $10.9 \pm 1.2$ yr
(Hathaway 2010), including the long cycle 4, which is the largest deviation
(e.g. the one and only $5~\sigma$ deviation for 17 yr from maximum to maximum).}
The Gnevyshev-Ohl rule was also fullfilled for those eight cycles,
in particular in sunspot group numbers.
Only after deviding the (otherwise very long) cycle no. 4 into two cycles,
the Gnevyshev-Ohl rule also holds for the following cycles until recently.

Schwabe cycles no. -2 and 3 were very intense,
and $^{14}$C decreased strongly in those years;
these are among the strongest decreases of $^{14}$C within 5-yr Intcal09
data listed in Table 2: AD 1725 to 1730 to 1735 to 1740
(partly in cycle no. -1) and AD 1775 to 1780 to 1785,
the one from 1780 to 1785 is even the strongest decrease between 1000 BC and AD 1900;
in addition, there was a strong decrease also AD 1760 to 1765 (part of cycle no. 1).

In cycles no. 3 and 4 (actually 4a), there was particularly intense auroral activity,
even during the sunspot minimum between cycles no. 3 and 4a
(for the sunspot minimum year AD 1784,
Fritz (1873) listed several aurorae 
in his southernmost zone 
in Padua, Italy);
the related solar wind helped $^{14}$C to stay low:
as we see in Fig. 2ab, in particular from AD 1770s to 1790s,
the evolution of auroral activity is reflected in the $^{14}$C curve.
Also, the evolution over those eight Schwabe cycles since the end of the
Maunder Grand Minimum (Fig. 3) clearly shows how well aurorae and $^{14}$C are anti-correlated --
not only in each Schwabe cycle -- also the $^{14}$C level decreases from cycle to
cycle while auroral activity increases.
While sunspots almost reach zero values in the sunspot minima,
the auroral wind level does not go down to zero in the minima,
but tend to increase from minimum to minimum.
This behaviour is seen until the early AD 1790s, when activity suddenly drops.

Interestingly, the recalibrations of sunspot numbers by Clette et al. (2014) 
and Svalgaard \& Schatten (2015) show that the activity in the 2nd half
of the 18th century was underestimated so far and was not weaker then in the
past few decades.

\subsubsection{Activity around AD 1795}

We will now discuss, whether and how the solar activity proxies
fit to the two suggested possibilities, with a long cycle no. 4
or after splitting no. 4 into the two cycles 4a and 4b
(see footnote 9, the sunspots alone cannot decide the issue, see Fig. 2de):
the sunspot minimum between cycles 4 and 5 at AD 1798.3 (dashed line in Fig. 2ef)
does not fit well with the aurorae (Fig. 2c) nor with $^{10}$Be (Fig. 2d),
while the increase in $^{14}$C from AD 1794 to 1798 (shifted
Stuiver et al. 1998 data, Fig. 2b) could be seen as to be due to the
decline of solar activity at the end of a long cycle no. 4;
however, the $^{14}$C data from Stuiver \& Braziunas (1998) rise almost continuously
from AD 1791 to 1796 (shifted data), but here only 2-yr time resolution.

If cycle 4 would be split into 4a and 4b (dotted lines in Fig. 2), with 4b to last until
a sunspot group minimum in AD 1799.8 (Usoskin et al. 2001a),
then it fits well the local aurora minimum around AD 1800 (Fig. 2c, for Tromholt
aurorae, we plot only those in his zones II to V, 
i.e. always south of the (typical) polar auroral oval),
as well as with a local $^{10}$Be maximum around AD 1800
(Fig. 2d shows that NGRIP and Dye-3 evolve very similar from AD 1794 to 1810).
Given the long strong drop in aurorae from AD 1789/90 to 1794/95 (Fig. 2ac),
the extra Schwabe cycle minimum may lie at around AD 1794/95,
where there is also a local $^{10}$Be maximum
(rather than in 1793.1 as in Usoskin et al. 2001a,
there was an aurora in Padua, Italy, in 
the southernmost zone 
of Fritz (1873) in Dec 1792).
Considering the small local aurora maximum around AD 1797/98 (Fig. 2ac),
as well as the sunspot minimum and in particular the $^{10}$Be maximum, both around AD 1800,
it might be the best compromise to place a Schwabe cycle minimum at around AD 1800
(rather than in AD 1798); cycle 4b would then have a length of some 5-6 yr.
The sunspot minimum around AD 1800 would be better consistent with the
$^{14}$C data of Stuiver et al. (1998),
if it would be shifted back by some 4 yr (instead of the usual 2.5 yr).\footnote{This 
effect seems to be present in the whole Dalton minimum.
It could be a result of a strong disturbance of the carbon cycle after around AD 1790
(shortly before, $^{14}$C data are missing for AD 1782 and 1783).}

From the proxies alone, it is hard to decide whether cycle 4 has to be split:
aurorae and $^{10}$Be show the additional minimum and maximum,
while the evidence from (poor) sunspots and (noisy) $^{14}$C is not that clear.

\subsubsection{Activity since around AD 1795}

After the very intense Schwabe cycle 4a, the short Dalton Minimum started,
which lasted for three to four Schwabe cycles:
no. 4b and 6 are particulary weak, while no. 7 already shows enhanced activity. 
At the start of the Dalton minimum, the auroral level droped strongly 
in the AD 1790s, but not down to zero (Fig. 2ac);
auroral activity all but vanished AD 1810-1815, so that $^{14}$C reached
its local maximum (Schwabe cycle no. 6). 
Afterwards, since AD 1825, auroral and sunspot activity rose, so that $^{14}$C droped.
At this time, i.e. at or after the end of the short Dalton Minimum, 
there was one of the six strongest decreases in $^{14}$C from AD 1830 to 1835 
in Intcal09 and Intcal13, and also the strongest decrease in SHCal13 in AD 1835-1840 (Table 2).

In the period AD 1795 to 1800, there was the strongest increase in $^{14}$C
(Table 1) in Intcal09, Intcal13, and SHCal13 5-yr time steps (from 1000 BC to AD 1900).
In SHCal13, $^{14}$C started to increase already in AD 1790 (unshifted raw data).
This increase was fast and strong as seen in data with higher time resolution:
for the northern hemisphere, in $^{14}$C data with 1 yr time resolution (Stuiver et al. 1998),
we can see an intense increase over four years from AD 1794 to 1798
(i.e shifted by 2.5 yr as plotted in Figs. 2, 3, and 10).
For the southern hemisphere, in $^{14}$C data with 2 yr time resolution (Stuiver \& Braziunas 1998)
we can see the strong increase already from AD 1791 (also shifted).

\begin{figure*}[t]
\vspace{-0.15cm}
\begin{center}
\includegraphics[width=10cm,angle=270]{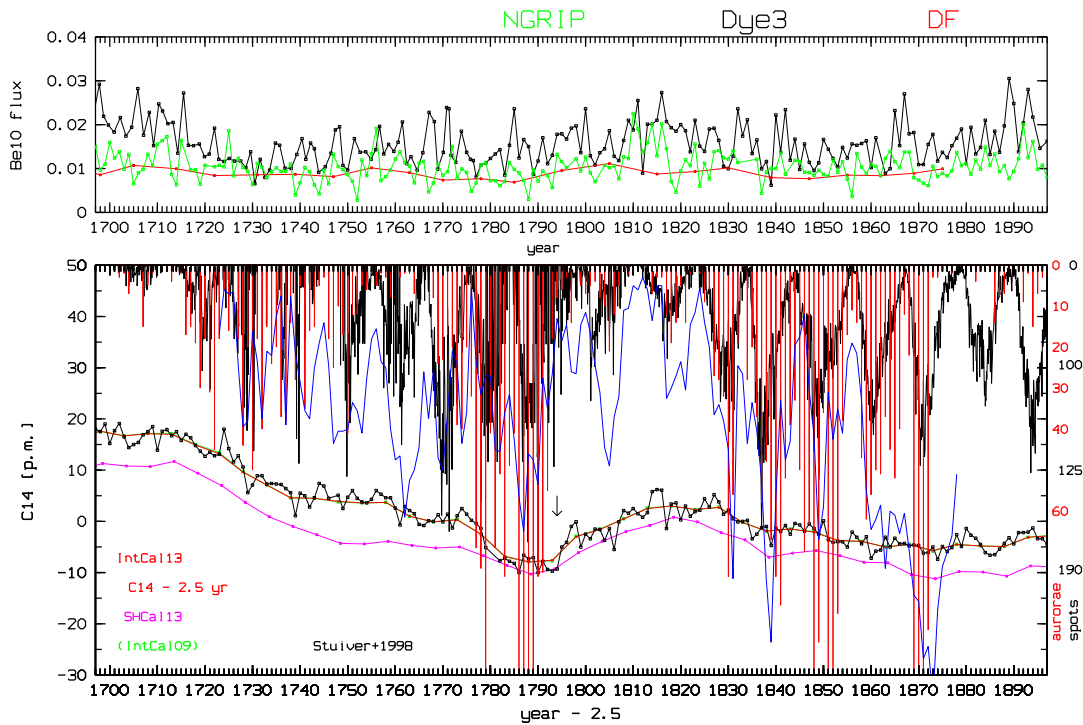}
\end{center}
\caption{{\bf Solar activity proxies AD 1697 to 1897:}
Top: $^{10}$Be flux in 0.01 atoms cm$^{-2}$ s$^{-1}$, Dome Fuji in red from Horiuchi et al. (2008),
NGRIP in green from Berggren et al. (2009), and Dye 3 in black from Beer et al. (1990).
Bottom: $^{14}$C (in p.m.) from Intcal09 (Reimer et al. 2009) in green
and Intcal13 (Reimer et al. 2013) in red mostly on top of Intcal09,
then yearly data from Stuiver et al. (1998) in black;
SHCal13 data (Hogg et al. 2013b) in pink.
Telescopic (monthly) sunspot group numbers from Hoyt \& Schatten (1998)
are plotted as black line pointing downward showing the Schwabe cycles.
Aurorae from Fritz (1873) as red lines and from Tromholt (1902) as blue line, also pointing downward,
consistent with the sunspot cycles.
For sunspots and aurorae, the scales are given at the bottom-right y-axis in black and red, respectively.
$^{14}$C data are plotted 2.5 yr {\em before} the measured time,
but spots, aurorae, and $^{10}$Be are plotted for their measurement year.
An arrow indicates the start of a strong increase in $^{14}$C.
Note the short Dalton minimum beginning with this increase.}
\end{figure*}

We can also see that $^{10}$Be flux is higher from AD 1794 to 1798
(mean: $0.012 \pm 0.001$ for NGRIP, $0.018 \pm 0.001$ for Dye 3, unit 0.01 atoms cm$^{-2}$ s$^{-1}$)
than from AD 1787 to 1793 (mean: $0.007 \pm 0.002$ for NGRIP, $0.013 \pm 0.003$ for Dye 3, 
unit 0.01 atoms cm$^{-2}$ s$^{-1}$), see Figs. 2 \& 3.
$^{10}$Be was also very low during cycle 3, when $^{14}$C has shown its largest decrease.
The increase of $^{10}$Be since AD 1793/94 was near the solar activity minimum
between 4a and 4b and the prolonged higher average also shows how weak cycle 4b 
(or the 2nd half of cycle no. 4) was.
In Dye-3 data, $^{10}$Be reached a similary high value AD 1810-1816 during Schwabe cycle 6 (Fig. 2).

Given that $^{10}$Be has lower absolute timing accuracy than $^{14}$C, 
we should also search for peaks {\em around} AD 1800.
Usoskin \& Kovaltsov (2012) found NGRIP peaks in AD 1810 and 1816 and a peak in Dome Fuji (DF) in 1805,
but wrote that {\em the 1805 AD peak in the DF series was rejected based on the annual NGRIP and Dye 3 data}.
We think that $^{10}$Be is dated well around 1800, as NGRIP and Dye 3 show their local
minimum $\sim 1790$, when solar activity is maximal. The high level $\sim 1810$ until 1816 
in Dye 3 and NGRIP was due to the fact that the rather weak Schwabe cycle no. 6 followed after the more
active cycle no. 5 and that the Schwabe cycle minimum inbetween 5 and 6 was very deep
(almost no aurorae).

\subsubsection{Summary on solar variations around AD 1795}

From the very strong decrease AD 1780 to 1785 over the strong increase
to the last strong decreases AD 1830 to 1840, we have a time span of 60 yr.
From the beginning of the Maunder Grand Minimum until AD 1840, we have roughly 200 yr.

A $^{14}$C increase due to a decrease in the relevant solar modulation potential
is caused not only by a decrease in visible sunspots and aurorae,
but by a decrease in the relevant solar wind (possibly mainly coronal hole wind);
this effect is also seen in the fact that the increasing solar activity (aurora)
level from activity minimum to minimum throughout the 18th century leads to a decrease of
the $^{14}$C level (otherwise, both the large Schwabe cycle amplitudes in solar
activity at the end of the 18th century and the low Schwabe cycle amplitudes in
solar activity in the Dalton minimum
result in similar Schwabe cycle $^{14}$C amplitudes).
Within the Dalton minimum, the $^{14}$C level
reaches its local maximum in the second activity minimum (around AD 1811, between cycles 5 and 6),
which may correspond to the solar modulation potential and wind floor (aurorae).

In summary, even though of the noise, we can essentially follow the evolution of solar activity
in $^{14}$C, which does react quite fast (within very few years) -- and even in $^{10}$Be;
the level of radioisotopes follows auroral activity (solar wind modulation).
The increase in $^{14}$C from AD 1790 to 1796 (or from 1794 to 1798) is then not (only)
due to the declining phase of a Schwabe cycle
(and the end of a Grand maximum),
but due to the sudden start of the Dalton Minimum,
i.e. a sudden strong drop also in the longer-term solar wind activity.
More $^{14}$C data with 1-yr time resolution would be taken
at around this time to clarify the situation.

\subsection{Solar activity variations around AD 775}

After several hundred years without any pronouced Grand Minima
and almost no intense rapid $^{14}$C variations (within 5 yr) in Intcal and SHCal 
(Tables 1 \& 2, Fig. 1),
there was the Dark Age Grand Minimum in the 7th century:
there were two strong rapid $^{14}$C increases within 5-yr in SHCal13 (Hogg et al. 2013b)
at AD 675 to 680 and to 685 (Table 1, Fig. 1).

We will discuss first the northern hemisphere data, then the southern data.

\subsubsection{The northern hemisphere data}

After the end of the $^{14}$C increase of the Dark Age Grand Minimum at the local Intcal maximum 
in the early AD 690s (according to data in Miyake et al. (2013b) with higher
time resolution, in AD 692 $\pm 2$), 
we have the following situation (Figs. 1 and 4):
there was strong activity for eight cycles (Grand Maximum):
$^{14}$C decreased until about AD 774 (shifted),
which we interpreted also as the ending minimum of a strong Schwabe cycle (Neuh\"auser \& Neuh\"auser 2015a).
\footnote{In Neuh\"auser \& Neuh\"auser (2015a), we dated all Schwabe cycle maximal and minimal
activity phases from AD 731 to 844 by taking into account aurorae, sunspots, and the $^{14}$C record.}
With the same method as in Neuh\"auser \& Neuh\"auser (2015a), 
where we already dated the four Schwabe cycles from AD 733 $\pm 1$ to about AD 774.
we dated four more Schwabe cycles from AD 692 $\pm 2$ until 733 $\pm 1$; 
the activity minima 
-- found as $^{14}$C maxima -- are displayed in Fig. 4 as upward pointing 
arrows.\footnote{One can also date phases of maximal and minimal activity, i.e. the Schwabe cycle, 
with $^{14}$C only: one first applies an 5-yr filter, i.e. one takes the average $^{14}$C
for 5-yr bins (the same time periods as bins as in Intcal), 
then one applies an 8-13 yr bandpass filter,
i.e. one compares the real $^{14}$C data with the 5-yr averages. 
The offsets between the real data
and the 5-yr averages then reflect the phases of maximal and minimal activity (see footnote 4).
To remove the long-term variations in $^{14}$C by averaging also has a disadvantage,
as it removes information -- namely on the current secular trend, e.g. decreasing or increasing wind level.
Given the time-resolution, the noise in $^{14}$C, and the carbon cycle, it is even better
to date the Schwabe cycles not only with $^{14}$C, but by also taking into account
aurorae and sunspots, as done in Neuh\"auser \& Neuh\"auser (2015a). Also, visual inspection
is very powerful for complex multi-proxy analysis.
Schwabe cycles modulations were found before for both $^{14}$C and $^{10}$Be data 
(e.g. Stuiver \& Quay 1980, Bard et al. 1997, Stuiver \& Braziunas 1998,
Nagaya et al. 2012) for the time since AD 1700,
and also for the Maunder Minimum, even though the Schwabe cycle amplitudes in radioisotopes
were somewhat smaller (Kocharov et al. 1995, Beer al. 1998, Damon et al. 1999, Usoskin et al. 2001b,
and Miyahara et al. 2005).}
To show that the $^{14}$C record around AD 800 really shows the typical Schwabe cycle
variation, we plot several Schwabe cycles on top of each other in Fig. 5.
In those eight cycles from AD 692 $\pm 2$ until about AD 774,
the mean cycle length was $10.9 \pm 2.3$ yr.

\begin{figure*}[t]
\vspace{-0.15cm}
\begin{center}
\includegraphics[width=10cm,angle=270]{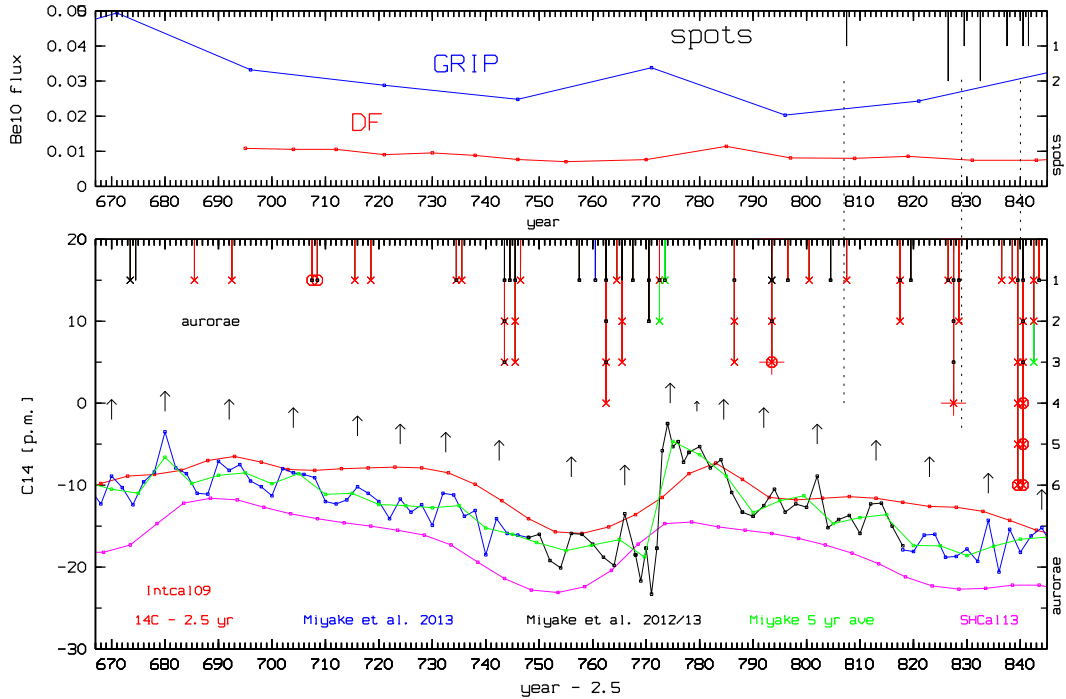}
\end{center}
\caption{{\bf Solar activity proxies AD 667 to 845:}
Top: $^{10}$Be flux as in Fig. 3 plus GRIP from Vonmoos et al. (2006) in blue;
naked-eye sunspots as black lines from the top from Vaquero et al. (2002) plus his updated (priv. comm.).
Bottom: $^{14}$C (in p.m.)
with 5-yr time reolution from Intcal09 (Reimer et al. 2009) in red,
SHCal13 (Hogg et al. 2013b) in pink,
and Miyake et al. (2013b) in blue, but the yearly averaged data from Miyake et al. (2012, 2013a)
in black for AD 750-820.
Aurorae from the publications cited in Sect. 3 are shown as follows:
blue, green, and red lines for aurorae with that colour, black lines for aurorae
where no color was reported, the length of the lines indicate the number of aurorae for each colour
in that year (scale at the bottom right edge),
correctly colored crosses (at the end of aurora lines) indicate aurorae where dynamics were reported,
circles indicate aurorae that lasted for several days (at least two days within
four consecutive days) possibly indicating CMEs,
the number of large plus signs indicate the number of aurorae, where even strong pulses were reported,
and the number of small black dots indicate the number of aurorae that year outside Europe
(Byzantium, Arabia, and Eastern Asia, i.e. stronger).
$^{14}$C data plotted 2.5 yr {\em before} the measured time,
but spots, aurorae, and $^{10}$Be are plotted for the measurement year.
The green curve shows the 5-yr averages of the (bi-)annual Miyake et al. data
(see Sect. 1 for details).
We identify all Schwabe cycle activity minima by upwards pointing arrows
from spots, aurorae, and $^{14}$C (the small arrow is more uncertain).
Dotted lines indicate phases of maximal activity ($^{14}$C minima)
in Schwabe cycles simultaneous to (clusters of) naked-eye sunspots.}
\end{figure*}

\begin{figure}[t]
\vspace{-0.15cm}
\begin{center}
\includegraphics[width=5cm,angle=270]{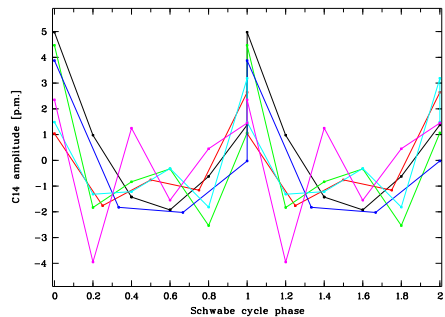}
\end{center}
\caption{{\bf Radiocarbon data of six Schwabe cycles plotted on top of each other:}
We selected the Miyake et al. $^{14}$C data from AD 784 to 844.
We separate the data into the six Schwabe cycles which all start and end with
the respective local $^{14}$C maximum.
We normalized those six cycles to the same length (being from 0.0 to 1.0).
Then, we plot the $^{14}$C data for those six
cycles versus the normalized (i.e. phased) time, i.e. on top of each other (in different colours).
We plot each cycles twice, i.e. from phase 0.0 to 2.0, for better visibility.
Even though of some noise, we can clearly see the modulation.
This shows that the Schwabe cycle variation can be seen in Miyake et al. $^{14}$C data.
}
\end{figure}

Note that (after strong activity, i.e. low $^{14}$C) the increase in $^{14}$C at AD 773-776 
is in the declining phase of a Schwabe cycle according to our reconstruction 
(see Fig. 3 and Neuh\"auser \& Neuh\"auser 2015a), followed by one or more weak cycle(s).
G\"uttler et al. (2015) also took into account the Schwabe cycle modulation 
accross the time span studied (but fixed the cycle length to 11 yr).
In their figure 2, one can see that the strong $^{14}$C increase at AD 774/5 is
in the declining phase (at the end) of a Schwabe cycle, i.e. consistent with our reconstruction.
Usoskin et al. (2013) used presumable aurorae to reconstruct the Schwabe cycles
in these decades and claimed to have found 
{\em a distinct cluster of aurorae between AD 770 and AD 776} suggesting 
{\em a high solar activity level around AD 775} plus {\em the next nearest
observations around AD 765-767 and AD 786, this suggests a 11-year cyclicity.}
Their {\em high solar activity level around AD 775} is thus in clear contradiction
to the reconstruction by G\"uttler et al. (2015) and us, that the $^{14}$C increase
was in the declining phase (or minimum) of a Schwabe cycle.
The presumable aurorae in the mid AD 770s (Usoskin et al. 2013) were 
most probably all halo displays (Neuh\"auser \& Neuh\"auser 2014, 2015ab, Chapman et al. 2015).

It is very unlikely to have a very strong super-flare in the declining phase (or minimum) of a Schwabe cycle:
none of the 30 strongest solar flares since 1978  
(www.spaceweather.com/solarflares/topflares.html) 
happened within two years before the minimum
(the closest of those 30 flares before a minimum was an X10 class flare on 2006 Dec 5, 
two years before the next minimum);\footnote{In addition, there was a (less severe) solar storm with a CME on 
1986 Feb 7-9 with $a_{p} \simeq 19$ and $k_{p} \ge 8$ a bit more than half a year before the minimum in 1986.8
(V. Bothmer, priv. comm.).}
most of the strong flares happen in and around the sunspot maximum
(also, the Carrington event in 1859 was years after the minimum in 1856.0).

The $^{14}$C data sets with 5-yr time resolution show several intense variations 
in the decades around AD 775 (see Tables 1 \& 2, Figs. 1 \& 4):
\begin{itemize}
\item We see three rather strong rapid decreases in $^{14}$C data at AD 735 to 740 to 745 to 750,
where the $^{14}$C decreased by 2.0 to 2.5 p.m. in a 5-yr step (small arrows in Fig. 1).
\item Shortly afterwards, there are strong rapid rises in $^{14}$C data at 
AD 765 to 770 to 775 (Intcal13, SHCal13) and from AD 775 to 780 (Intcal09).
\item Finally, we have two somewhat weaker rapid decreases in $^{14}$C data at AD 785 to 790 to 795 (Intcal),
where the $^{14}$C decreased by 2.0 to 2.5 p.m. in a 5-yr step.
\end{itemize}
From the first to the last strong rapid decrease (AD 735-740 to AD 790-795), we have 60 yr.
From the beginning of the Dark Age Grand Minimum (beginning of 7th century) until the last 
strong rapid decrease (AD 790-795), we have less than 200 yr of rapid $^{14}$C evolution.

Auroral activity (indicating solar wind) was relatively intense around AD 745 
and even more so AD 757-773,\footnote{Strong
aurorae were indicated by color, dynamics, repetition, simultaneous sightings
in different areas, and low geomagnetic latitude, see Neuh\"auser \& Neuh\"auser (2015a)
for details. Some particulary strong aurorae, e.g. at AD 762, 793, and around 840
(stronger than in the 770s), did not lead to additional $^{14}$C peaks.}
and $^{14}$C decreased substantially from the beginning of the AD 740s until the early 770s.
The two Schwabe cycles just before AD 774 both lasted some 8-9 yr and were very strong
(the third-to-last cycle before AD 774 was also intense), 
clearly seen in aurorae and $^{14}$C, Fig. 4.
From about AD 774 to 784, we can fit in either one average-length or two short weak Schwabe cycles
(weak due to the lack of aurorae and high $^{14}$C from AD 774 to 785);
after AD 784, the Schwabe cycles are well visible in $^{14}$C (larger amplitudes),
aurorae, and some naked-eye sunspots 
(see Figs. 4 and 5 and Neuh\"auser \& Neuh\"auser 2015a for details).

\subsubsection{The southern hemisphere data}

In Fig. 6, we can see that the $^{14}$C rises around AD 775 and 994 are detected also on the
southern hemisphere. SHCal13 is based for those centuries on two data sets, namely
Zimmerman et al. (2010) for Tasmania Huon and Hogg et al. (2011) for New Zealand Kauri,
all plotted separately in Fig. 6. These trees show the sharp rise in $^{14}$C, namely 
from $-20$ p.m. in AD 765 to $-9$ p.m. in 775 in data by Zimmerman et al. (2010)
and from $-24$ p.m. in AD 765 to $-12$ p.m. in AD 775 in data by Hogg et al. (2011).
Zimmerman et al. (2010) noticed that the otherwise usual interhemispheric offset in $^{14}$C
(in p.m. or $^{14}$C age, Fig. 6a) was near zero from about AD 775 on for a few decades
in their data, which was different in Hogg et al. (2011).
Therefore, Hogg et al. (2013a) re-measured data from the same Huon tree as used in Zimmerman et al. (2010) 
for the decades around AD 800 and could find the usual interhemispheric offset
(i.e. could not confirm Zimmerman et al. 2010), an example for inter-laboratory offset; 
in the revised Huon data by Hogg et al. (2013a),
the rise is from $-24$ p.m. in AD 755 to $-14$ p.m. in both AD 775 and 785;
around AD 850, the data by Hogg et al. (2013a) are not consistent with those in Hogg et al. (2011).
There is a very large and sharp increase to AD 775 seen in all these data sets.
In both Zimmerman et al. (2010) and Hogg et al. (2011), the rise around AD 993/4 is also seen.
Also at AD 995, the single data point by Zimmerman et al. (2010) does not show
the otherwise usual interhemispheric offset in $^{14}$C, contrary to Hogg et al. (2011, 2013b). 
Interestingly, the interhemispheric offset seems to be small to vanishing at the
large sharp $^{14}$C rises around AD 775, 995, and 1795 (Fig. 6a) -- possibly due to a disturbance
in the south ocean wind strength, which is otherwise thought to be responsible for the
hemispheric offset; there are a few more such cases seen in Fig. 6a.

\begin{figure*}[t]
\vspace{-0.15cm}
\begin{center}
\includegraphics[width=10cm,angle=270]{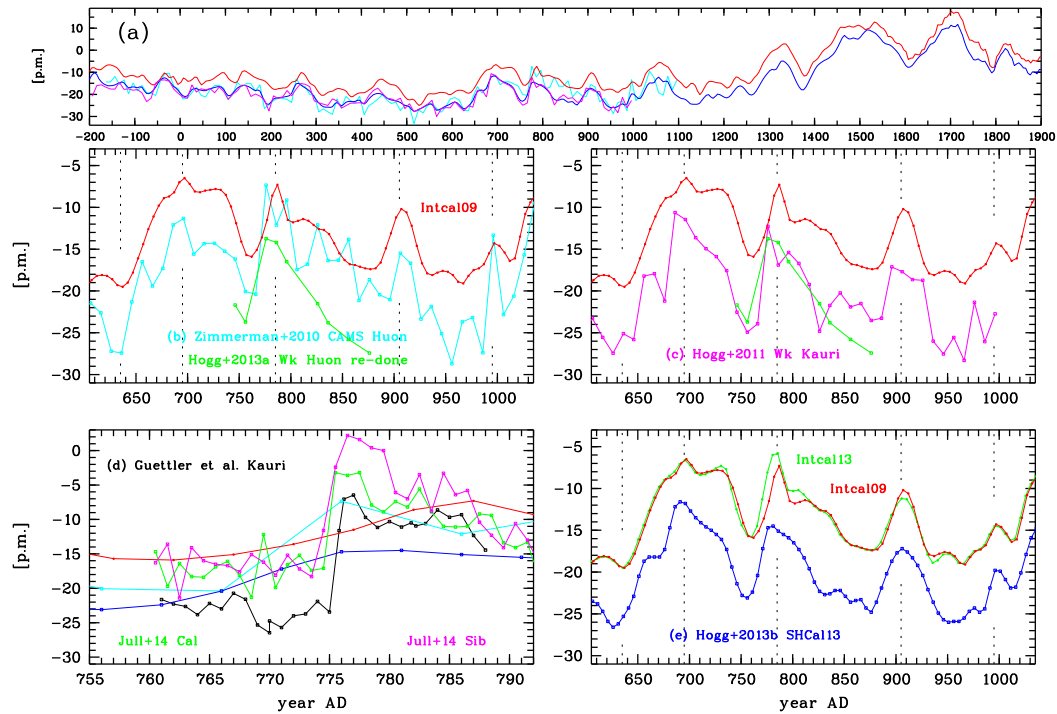}
\end{center}
\caption{{\bf Radiocarbon $^{14}$C for the southern hemisphere:}
(a) $^{14}$C from BC 201 to AD 1900 from Intcal09 (Reimer et al. 2009) in red for the northern hemisphere
compared to data from the southern hemisphere from SHCal13 (Hogg et al. 2013b) in dark blue,
Zimmerman et al. (2010) in light blue, and Hogg et al. (2011) in pink showing the known
interhemispheric offset for most of the period; Intcal and SHCal with 5-yr time resolution,
the other sets with 10-yr. The data are plotted in the same colours in the other plots, too,
unless otherwise noted.
(b) CAMS data by Zimmerman et al. (2010) for a Huon tree compared to Intcal09
showing that there is almost no interhemispheric offset around AD 800,
but a strong fast rise from AD 765 to 775.
Wk data from Hogg et al. (2013a) in green for the same Huon tree measured by Zimmerman et al. (2010),
showing the interhemispheric offset (and the strong fast rise from AD 755 to 775).
(c) Wk data from Hogg et al. (2011) in pink for a Kauri tree and
Wk data from Hogg et al. (2013a) for the same Huon tree measured by Zimmerman et al. (2010),
compared to Intcal09.
(d) Data with 1-yr time-resolution from G\"uttler et al. (2015) for a southern Kauri tree
in black compared to the data with 1-yr time-resolution from Jull et al. (2014) for
the Siberian (pink) and Californian tree (green),
as well as to the Huon tree measured by Zimmerman et al. (2010),
Intcal09, and SHCal13.
The northern data from Jull et al. (2014) are plotted into the years x.5, as the tree rings were
formed mainly in northern spring and summer, i.e. in the middle of the calendar year,
while the southern data from G\"uttler et al. (2015) are plotted into the years x.0, as the
tree rings were formed mainly in the southern spring and summer, i.e. around the turn of the
calendar year; note that a tree ring labeled e.g. 774 in G\"uttler et al. (2015) was formed in
775.0 as specified by G\"uttler et al. (2015); (also note that the tree rings labeled 775, 775.5,
781, and 781.5 were formed in (and are plotted into)
775.8 (Sep-Dec), 775.2 (Dec-Mar), 781.8 (Sep-Dec), 781.2 (Dec-Mar) according to L. Wacker, priv. comm.).
The data by Jull et al. and G\"uttler et al. do not show any calendar year offset or time-shift,
but an increase in radiocarbon from AD 773.5 (Jull et al.) or 775.0 (G\"uttler et al.) until
AD 779.5 (Jull et al.) or 777.0 (G\"uttler et al.). Data are not shifted by 2.5 yr in these plots.
(e) SHCal13 (Hogg et al. 2013b) compared to Intcal09 (Reimer et al. 2009)
and Intcal13 (Reimer et al. 2013) in green, all with 5-yr time steps.
All southern data sets show the strong fast rise to AD 775 and AD 995 (and others).
The dotted lines indicate local maxima and minima in the data sets, showing that there is
an calendar year offset before about AD 800 by a few yr, also seen in panel (a).}
\end{figure*}

Hogg et al. (2013b) then use the revised Huon tree data from Hogg et al. (2013a) for SHCal13.
The increase from AD 775-780 is neither
that large nor that significant (in SHCal13 compared to Intcal), 
while the increases in AD 765-770 and 770-775 are strong (Table 1).
In Fig. 6, we see that the $^{14}$C variation started earlier in SHCal13.
In Fig. 1, we see a calendar age offset between Intcal and SHCal13 from
the 2nd century BC to the ninth century AD: SHCal13 appears to be some 5 yr earlier than Intcal.
Before BC 200, SHCal13 data are essentially identical to Intcal13 except the interhemispheric offset
in the $^{14}$C content (Hogg et al. 2013b).
This offset in calendar age can probably not be caused only 
by carbon cycle differences between the northern and southern hemispheres,
because we see the offset (by some 5 yr) only from roughly BC 200 to AD 800 (Fig. 6a);
a calendar age offset between the northern and southern hemisphere due to different growing seasons
can amount to only half a year.
Furthermore, the decrease after AD 775 took longer than expected for a short one-year event (Fig. 6e).

If the increases in $^{14}$C would have started in different years on the two different hemispheres,
this could speak against a solar origin, but for an extra-solar origin. If the event started earlier
on the southern hemisphere, one would conclude on an origin on the southern sky: because the Sun is currently
above the Galactic plane, supernovae from the Galactic disk would be expected preferentially on
the southern sky. The same would be expected for Short Gamma-Ray Bursts, in particular also,
because the Galactic Center with largest event rate is located on the southern sky.
Stratospheric ionization due to a Short Gamma-Ray Burst (few seconds) would deplete the local ozone
layer due to NO$_{\rm x}$ mainly around the sub-{\em stellar} point; this leads to a drop in
stratospheric temperature and a rise in tropopause altitude causing larger air mass exchange
and injection of radionuclids from the stratosphere into the troposphere (e.g. Pavlov et al. 2014).
Such effects may effectively disturb in the otherwise normal interhemisperic offset.

However, we do not claim that the hemispheric offset in calendar age (in data with 5-yr resolution) points
to an extra-solar origin. Instead, either the northern Intcal09 could be a few yr too late
(less in Intcal13) and/or the southern SHCal could be a few yr too early. 
By comparison of both Intcal and SHCal13 with the 1- to 2-yr time resolution data, 
one could conclude that both northern Intcal and southern SHCal are off by very few years. 

G\"uttler et al. (2015) then took data with one year time resolution for a southern Kauri tree
and found that the strong increase in $^{14}$C happened from AD 775.0 ($^{14}$C produced
in the southern summer of 774/5 deposited in the tree ring labeled 774 in table 3 in
G\"uttler et al. 2015) to 776.0.
According to the previously known northern data with one year time resolution, the increase
in $^{14}$C happened after the incorporation of $^{14}$C in the northern summer of 774, 
i.e. after around 774.5, but before around 775.5. Northern and southern data together
constrain the timing of the strongest yearly increase to have 
started between AD 775.0 and 775.5 (unshifted).\footnote{This
conclusion can be drawn only under the assumption that the five northern trees and the one
southern tree with one year time resolution data all have experienced exactly the same number of
years with zero tree-growth, which may be possible, e.g. due to excessive coldness, see e.g. Mann et al. (2012);
the number of such years may or may not be zero.
Also note that $^{14}$C started to increase already in AD 773 in several data sets.}
Solar activity was already decreasing since AD 772/3 (unshifted).
This is well consistent with the considerations and the sequence of events (last reported 
aurora curtains near
the Turkish-Syrian border at AD 772 and 773) as discussed in Neuh\"auser \& Neuh\"auser (2015a).
The SHCal13 data stay roughly constant at $-15$ p.m. for several decades since AD 775.

We have binned the G\"uttler et al. (2015) data into the same 5-yr-bins as SHCal13, see Fig. 7, where we can see
that their data are not consistent with SHCal13 (Hogg et al. 2013b), except the first bin,
and also deviant from different Kauri tree(s) from Hogg et al. (2011), which were used to generate SHCal13
(such a difference can be due to offsets between different laboratories and different trees,
A. Hogg \& L. Wacker, priv. comm.).
Interestingly, the G\"uttler et al. (2015) data are consistent with the lower resolution Zimmerman et al. (2010)
data, see Figs. 6 \& 7. Hence, neither the G\"uttler et al. (2015) nor the Zimmerman et al. (2010) data
show the interhemisperic offset (in $^{14}$C age or p.m.) just around AD 775, which is otherwise
mostly seen. 
We can also see in Fig. 6, that the strong increase in $^{14}$C from AD 765 to 775 in Hogg et al. (2011)
with 10-yr time resolution is not that strong anymore in SHCal13 (Hogg et al. 2013b) with 5-yr time
resolution, even though the latter are derived from the former; the SHCal13 interpolation proceedure
seems to take away strong slopes.

\begin{figure}
\vspace{-0.15cm}
\begin{center}
\includegraphics[width=5.5cm,angle=270]{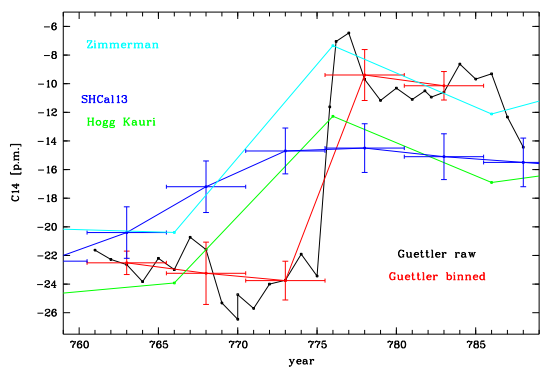}
\end{center}
\caption{{\bf Radiocarbon $^{14}$C for the southern hemisphere around AD 775:}
Raw data from G\"uttler et al. (2015) in black,
and binned into 5-yr bins in red (same bins as in SHCal13, and in the three cases, where there
were two data points in G\"uttler et al. (2015) within less than one year, those two data points
where averaged first, so that we had only one data point per year),
SHCal13 data in dark blue (Hogg et al. 2013b), 
the Kauri tree data from Hogg et al. (2011) with 10-yr time resolution, 
on which SHCal13 is based here, in green,
and the southern tree from Zimmerman et al. (2010) in light blue.
One can see that the G\"uttler et al. (2015) data, binned into 5-yr-bins, 
are deviant from Hogg et al. (2013b) by more than $1 \sigma$ except in the first bin.
The G\"uttler et al. (2015) data are, however, consistent with the Zimmerman et al. (2010) data.
SHCal13 is made from data in Hogg et al. (2013a) in this epoch, but has a reduced slope in the increase.
}
\end{figure}

\subsection{Solar activity variations around BC 671}

After several centuries of relatively constant solar activity without Grand Minima
(Fig. 1), a Grand Minimum started
slightly before a strong rapid rise in $^{14}$C from BC 801-796 in Intcal, see Fig. 1
(SHCal13 does not provide additional information for this epoch); 
this Grand Minimum had its local maximum in $^{14}$C at BC 746-726.
Then, we have the typical sequence of strong decreases: BC 696-691,
BC 691-686 (the second-most strongest decrease in a 5-yr step),
BC 686-681, then the strong rise BC 671-666,
and then again a strong decrease BC 651-646 (all in Intcal), separated by 50 yr,
ending less than 200 yr after the start of the preceeding Grand Minimum. 
A strong increase from BC 676-656 was listed in M12, but not in Usoskin \& Kovaltsov (2012).
The period is part of the well-known Hallstadt plateau (in $^{14}$C age).

Takahashi et al. (2007) have taken 30 data with higher resolution for a Japanese cedar tree
with 1 to 10 yr time resolution -- the tree was covered by volcanic ash some 2500 yr ago.
According to Takahashi et al. (2007), their data range from 2757 to 2437 BP
and the outer edge tree ring has a calendar age of $2427.5 \pm 12.5$ BP,
i.e. a dating uncertainty of $\pm 12.5$ yr ($95\%$ confidence level).

The data by Takahashi et al. (2007) have a 19-yr calendar age offset (best fit) from Intcal,
which is also seen in their figure 2; 
we therefore shift their data by +19 yr, see our Fig. 8 (where we otherwise plot unshifted data);
their data then cover a range from BC 731 to 457.
The highest time resolution is from BC 685 to 621 (after having shifted by those 19 yr),
where the eleven data points taken all have a time resolution of 1 yr.
This period covers the strong variations listed above, 
but with coarse coverage -- with the strong decrease having the best coverage.

From BC 746 (start of local $^{14}$C maximum) to 671 (start of the strong rise in a 5-yr step),
we have $\sim 75$ yr, so that there could well be eight strong Schwabe cycles
(Grand Maximum). 
Some of those Schwabe cycle modulations are seen in the data by Takahashi et al. (2007),
in particular from BC 726 to 686 (data shifted by 19 yr).
Here, the amplitude of the Schwabe cycle modulation is $\sim 4$ p.m. (peak-to-peak),
i.e. intermediate between Miyake et al. (2013ab) and Stuiver et al. (1998) data (see Sect. 5.2).

The whole variation resembles well the situation from the end of the Maunder grand minimum
to the short Dalton minimum -- in particular those $\sim 20$ yr from
the strong decrease in $^{14}$C from BC 691-686 (the 2nd strongest in Table 2,
i.e. very intense activity, maybe the 2nd most intense activity in the three millenia before AD 1900)
to the strong rise BC 671-666 (Figs. 1 \& 8). 

\begin{figure}
\begin{center}
{\includegraphics[angle=270,width=8cm]{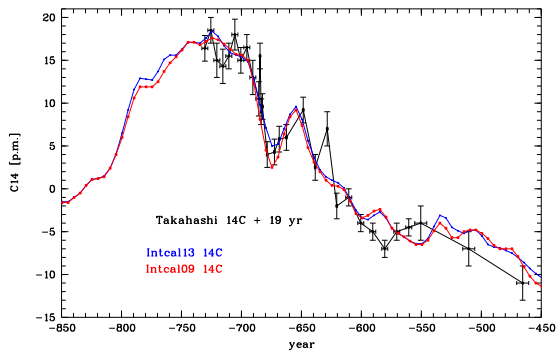}}
\caption{{\bf Variation of $^{14}$C from BC 851 to 451:}
$^{14}$C from Takahashi et al. (2007) in black 
(shifted by +19 yr to fit with Intcal),
Intcal09 and Intcal13 data from Reimer et al. (2009, 2013) in blue and red (unshifted).
There is a strong fast increase from BC 671,
but the best time resolution in Takahashi et al. (2007) is for earlier calendar years.}
\end{center}
\end{figure}

\section{Discussion: Strong rapid drop in solar activity from a high level}

We will first summarize the similarities between the cases discussed above
and discuss the three effects that caused the strong $^{14}$C variations. 
Then, based on the $^{14}$C rises in terms of p.m. in the two data sets
around AD 775 and 1795 and a comparison of the typical amplitudes in those
data sets, we can show that they have a similar order-of-magnitude.
Then, we consider whether such a sharp strong drop in the solar modulation
potential and the carbon production rate is possible quantitatively.

\subsection{Similarities in BC 670s/660s, AD 770s, and AD 1790s}

The three cases discussed, around BC 671, AD 775, and AD 1795,
are quite similar in secular behaviour of solar activity:
\begin{enumerate}
\item First, a grand minimum (in the 8th century BC,
then the Dark Age, and finally the Maunder Grand Minimum, respectively), then
\item a phase of some eight Schwabe cycles with increasingly
strong activity (Grand Maxima) with one or a few 5-yr-steps,
where $^{14}$C decreased by at least 2.5 p.m. (Fig. 1, Table 2) 
leading to a few cycles with very intense activity
(e.g. in the AD 760s and the 1780s).
\item Then, $^{14}$C suddenly increased by at least 2.5 p.m. over 5 yr
within one or two 5-yr steps (Table 1),
e.g. in the AD 770s and the 1790s.
\end{enumerate}
For the turn to the 19th century, this marked the start of the short Dalton minimum,
which lasted three to four Schwabe cycles.
Also, after the AD 775 $^{14}$C variation, solar actvity remained low
(no aurorae observed for more than a decade, Neuh\"auser \& Neuh\"auser 2015a).
In all those three cases, we have a series of 
\begin{enumerate}
\item strong rapid $^{14}$C drops,
\item rises, and 
\item drops within 60 yr each 
\end{enumerate}
(the final drops end less than 200 yr after the
start of the preceding Grand Minimum).

For a full understanding of the rapid strong $^{14}$C rise (like in the AD 770s and 1790s),
it is important to also consider the strong rapid decrease a few decades earlier.
We see three effects causing for the $^{14}$C rises (best seen around AD 1790s):
\begin{enumerate}
\item Auroral activity (indicating solar wind) remained very intense for at least 10-20 yr
(even during the activity minimum) before the rapid strong $^{14}$C rise,
so that $^{14}$C had dropped significantly to a very low level. 
\item In the declining part of the last strong Schwabe cycle,
when auroral and sunspot activity all but ceased, $^{14}$C rose up from
a very low level -- and on a short time-scale. 
\item After the rapid strong $^{14}$C rise, solar activity remained very weak for
up to four cycles, so that $^{14}$C kept increasing (AD 1790s) or stayed high (AD 770s).
\end{enumerate}

If weak activity follows shortly after a period of strong activity, 
solar wind density and/or velocity is also droping suddenly (not neccessarily down to zero),
leading to a strong and rapid increase in cosmic ray flux and radioisotope production;
one could speculate that such a $^{14}$C spike is particulary strong,
when the following Schwabe cycle is very weak and short, as the lost cycle in the AD 1790s;
this may have happened also in the 770s.

We still have to consider whether such a strong rapid drop 
in solar activity and solar modulation potential --
as an essential part of the secular variation seen within 200 yr --
is possible quantitatively.

\subsection{$^{14}$C increases in terms of p.m. and solar modulation drops}

Let us first consider the $^{14}$C increases in p.m. 
and the corresponding solar modulation potential decrease for the two best
studied periods under discussion
(all values and years for unshifted $^{14}$C data).

AD 1790s: $^{14}$C increased by $+4.8 \pm 1.4$ p.m. from AD 1795 to 1800 (Intcal09),
and from $-9.3 \pm 1.80$ p.m. in AD 1796 to $-5.0 \pm 1.70$ p.m. in AD 1797 and further to $-0.1 \pm 1.70$ in AD 1800,
i.e. it increased by $+4.3 \pm 2.5$ p.m. in 1 yr or by $+9.2 \pm 2.5$ p.m. in 4 yr 
(in data with 1-yr time resolution from Stuiver et al. 1998);
for the southern hemisphere, $^{14}$C increased from (the local minimum) -14.1 p.m. in AD 1793
to -9.0 p.m. in AD 1795 and further up to -11 p.m. in AD 1797 and then to -2.7 p.m. in the local maximum in AD 1803
(in unshifted data with 2 yr time resolution), the largest rise in a 2-yr time-step is +5.1 p.m.,
i.e. consistent with the northern hemisphere, but apparently starting slightly earlier 
(Stuiver \& Braziunas 1998). 

The solar modulation potential has been estimated for this time before,
based on solar activity proxies with only 10-yr time steps:
in the decade from AD 1795 to 1805, there is the strongest decrease
in the solar modulation potential as reconstructed
from (decadel) sunspots by Usoskin et al. (2002), from $^{14}$C by Solanki et al. (2004),
and from $^{10}$Be by Usoskin et al. (2003) and McCracken et al. (2004) --
while the strongest drop the (decadel) sunspot reconstructions by Solanki et al. (2004) 
and Usoskin et al. (2014) is in the decade AD 1785-1795, a little more than in the next decade;
also, according to Muscheler et al. (2005, 2007), 
the strongest decadel decrease in the solar modulation potential 
was in the decade from AD 1785 to 1795 from $\sim 1100$ MV to $\sim 650$ MV (Muscheler et al. 2005)
or from $\sim 900$ MV to $\sim 450$ MV (Muscheler et al. 2007),
the strongest decadel drop since the calculation start in AD 1500. 
From the $^{10}$Be increase in the 1790s, the respective solar modulation potential was estimated to have
decreased from $\sim 500$ MV in AD 1795 to $\sim 200$ MV in 1800 (Usoskin et al. 2003)
or from $\sim 450$ MV in AD 1795 to $\sim 250$ MV in 1800 (McCracken et al. 2004). 
Most recentlz, Usoskin et al. (2015) have shown that the number of auorae, 
the solar open source flux, the solar modulation potential, and also the
total solar irradiance all dropped suddenly and fast 
in the 1790s (their figures 10-13 and 17), namely at the sudden start
of the Dalton minimum, and that also the $^{14}$C production rate and the $^{10}$Be
flux increased from a very small to a relatively high value 
in the 2nd half of the 18th century (their figures 14 \& 15).

While this strong drop in solar modulation potential was calculated for a whole decade,
$^{14}$C increased strongly in an even shorter time interval: while Intcal09 data increased
by $+4.8 \pm 1.4$ p.m. from AD 1795 to 1800 (the largest rise in a 5-yr-step),
$^{14}$C did increase by $+4.3 \pm 2.5$ p.m. in the year from AD 1796 to 1797 (see Figs. 2, 3, 10).
Hence, most of the strong drop in solar modulation potential was in one to few years. 
This was observed in the AD 1790s at the turn to the short Dalton minimum. 

AD 770s: $^{14}$C increased by $+2.9 \pm 2.3$ p.m. in the 5-yr-interval AD 775-780 (Intcal09),
or from $-23.3 \pm 2.8$ p.m. in AD 773 to $-2.5 \pm 1.5$ p.m. in AD 776 
(and from $-17.7 \pm 1.5$ in AD 774 to $-5.8 \pm 1.8$ p.m. in AD 775
as the largest increase in a 1-yr time step), all from the M12 data averaged for their two trees,
i.e. it increased by $+11.9 \pm 2.3$ p.m. in 1 yr or by $+20.8 \pm 3.2$ p.m. in 3 yr (similar for other trees).
The largest increases within 1 yr ($+11.9 \pm 2.3$ p.m. AD 770s 
and $+4.3 \pm 2.5$ p.m. AD 1790s) are less than $2~\sigma$ different from each other.
The total increase in the AD 770s by $+20.8 \pm 3.2$ p.m. in 3 yr 
(M12 data) is only $2~\sigma$ different from the total increase in the
1790s by $+9.2 \pm 2.5$ p.m. in 4 yr. 
For the Siberian Larch in Jull et al. (2014), $^{14}$C in AD 775 is less than
$2~\sigma$ higher than in AD 774; in the Mannheim lab data of Usoskin et al. (2013),
$^{14}$C increased for 4 yr. 

When comparing the M12 increase in $^{14}$C with 1-yr time steps in the AD 770s 
with the Stuiver et al. (1998) 1-yr time steps in the AD 1790s,
one should also take into account that the amplitude of $^{14}$C variations
on short time-scales (Schwabe cycle) in M12 is on average larger than in Stuiver et al. (1998):
the $^{14}$C Schwabe cycle half-amplitude in Miyake et al. data (see Sect. 1) is $2.4 \pm 1.2$ p.m., 
while it only is $1.68 \pm 0.83$ p.m. in Stuiver et al. (1998) 
data; i.e. $42~\%$ more in M12 than in Stuiver et al. (1998).\footnote{both measured in the same way as outlined in Sect. 1,
but using only either odd or even years for Stuiver et al. (1998) data to simulate a bi-annual time-resolution as in
Miyake et al. data; for Stuiver et al. (1998) data, we covered the period AD 1700 - 1950
and compared the real data with the 5-yr-average.}
Hence, the amplitudes in Miyake et al, data are somewhat larger then in Stuiver et al. data, 
so that the increase in $^{14}$C by $+11.9 \pm 2.3$ p.m. in 1 yr or by $+20.8 \pm 3.2$ p.m. in 3 yr in M12 data
is not twice as large as the increase by $+4.3 \pm 2.5$ p.m. in 1 yr or by $+9.2 \pm 2.5$ p.m. in 4 yr in the AD 1790s 
in Stuiver et al. (1998) data.
We show in Figs. 9 \& 10, that at least the two $^{14}$C increases in AD 770s and 1790s 
lasted for several years (using $^{14}$C data with 1- to 2-yr time resolution),
not just for one year, fully consistent with strong decreases in solar modulation potential for 3-4 yr.

As mentioned in Sect. 2, there are some other 
strong short-term rises in $^{14}$C data with high time resolution (all years unshifted here):
\begin{itemize}
\item at AD 762/763 an increase by $+7.3 \pm 4.9$ p.m. (Jull et al. 2014 Siberian tree),
\item at AD 768/769 by by $+9.1 \pm 5.0$ p.m. (Jull et al. 2014 Californian tree),
\item at AD 792/793 by $+7.9 \pm 4.9$ p.m. (Jull et al. 2014 Siberian tree),
\item at AD 993-994 by $+9.2 \pm 2.6$ p.m. (Miyake et al. 2013a),
somewhat weaker in Menjo et al. (2005), probably also due to declining activity 
in a Schwabe cycle (and a weak cycle afterwards),
\item AD 1009-1010 by $+3.5 \pm 2.8$ p.m. (Menjo et al. 2005),
who argued that the strong rapid increase could be due to Schwabe cycle modulation,
\item at AD 1009-1011 by $+12.2 \pm 5.2$ p.m. (Damon \& Peristykh 2000),
who considered whether that was due to SN 1006,
\item at AD 1149-1150 by $+5.5$ p.m. (Damon et al. 1995, 1998, Damon \& Peristykh 2000),
\item at AD 1808-1809 by $+5.4 \pm 2.5$ p.m. (Stuiver et al. (1998).
\end{itemize}
These strong increases within 1-2 yr each may well be
a consequence mainly of a short strong declining phase in solar activity at the end of a Schwabe cycle.
The cases just listed are, however, not part of a strong secular variation (strong drops, rises, and
drops in $^{14}$C
after a Grand Minimum), i.e. somewhat different to the cases in the AD 770s and 1790s.

If the $^{14}$C increase in the 1790s corresponded to a decrease in solar modulation potential 
by $\sim 450$ MV in 5 yr, then the $^{14}$C increase in the 770s 
would correspond to a decrease in solar modulation potential by $\sim 900$ MV within few yr
(if twice as large), or less after the normalization 
due to different amplitudes in M12 and Stuiver et al. (1998);
it decreased strongly in a short time, but not neccessarily down to zero:
around AD 1800, some aurorae were still detected (e.g. Krivsky \& Pejml 1988), 
so that solar activity had not reached a zero level.
According to Poluianov et al. (2014), the solar modulation potential is around 1000 MV
in a Schwabe cycle maximum during high-activity phases and around 80 MV in the
Schwabe cycle minimum during low-activity phases like Grand Minima.
Now, when very weak activity follows shortly after very strong activity, like at the
sudden start of the Dalton minimum, one would expect a short-term drop in
solar modulation potential from $\sim 1000$ MV to $\sim 80$ MV.

It would be useful to reconstruct the $^{14}$C production rate,
the solar modulation potential, and the Total Solar Irradiance 
quantitatively around the AD 770s and the 1790s from $^{14}$C data with 1-yr time resolution, 
but this would be beyond the scope of this paper.

$^{10}$Be also increased in the decade AD 775-785 (M12).
According to Beer et al. (2013), the $^{10}$Be production rate would increase by a factor of a few,
when the solar modulation potential decreases from $\sim 1000$ MV to $\sim 100$ MV. 

While Solanki et al. (2004), Muscheler et al. (2007), and Steinhilber et al. (2012)
argue that the solar modulation potential as reconstructed from $^{14}$C and/or $^{10}$Be
can decrease to (nearly) zero (e.g. in the Maunder Grand Minimum, figure 12 in Usoskin 2013)
and also down (solar modulation potential and solar wind B) to 
(nearly) zero in the 7th/8th century AD (Steinhilber et al. 2012, Cliver et al. 2013 figure 13),
we do neither claim nor exclude that the solar modulation potential decreased to zero,
but -- as discussed above -- it decreased from a very large ($\sim 900$ MV) 
to a small value (hundred(s) MV) within a few years.

One could speculate that the effect, a strong rapid drop in solar activity and, hence, a 
sudden rise in $^{14}$C, can happen after a few decades of particular strong solar activity,
as just before the AD 1790s (maybe also before AD 775 and before BC 671).

\subsection{$^{14}$C production rate}

The (average background) atmospheric $^{14}$C production rate was estimated to lie at
$\sim 2.8$ to $2.50 \pm 0.50$ $^{14}$C atoms cm$^{-2}$ s$^{-1}$ (Lal 1988, Lingenfelter 1963)
or at $\sim 1.88$ to $1.64$ $^{14}$C atoms cm$^{-2}$ s$^{-1}$ (Kovaltsov et al. 2012),
depending on the strength of the geomagnetic field, the solar modulation potential, and 
numerical assumptions. The mean $^{14}$C (background) production rate
was also found to varry from $\sim 1.2$ during the 1990 solar maximum 
to $\sim 2.2$ atoms cm$^{-2}$ s$^{-1}$ during the 2010 solar minimum (Kovaltsov et al. 2012). 

For the $^{14}$C spike around AD 775, Usoskin et al. (2013) found an increase in the $^{14}$C production rate
of 1.1 to $1.5 \cdot 10^{18}$ atoms cm$^{-2}$, i.e. $4 \pm 1$ $^{14}$C atoms cm$^{-2}$ s$^{-1}$
(Usoskin et al. 2013, Pavlov et al. 2013) -- in addition to the background at that time 
(assumed to be 1.6 atoms cm$^{-2}$ s$^{-1}$ in Usoskin et al. 2013) --
estimated for an assumed {\em event} duration of {\em one} year only;
the value is smaller when considering that $^{14}$C may have been produced for several years,
which is also consistent with the increase for at least 3 yr (both AD 770s and 1790s),
e.g. it would be $\sim 1.3 \pm 0.3$ additional $^{14}$C atoms cm$^{-2}$ s$^{-1}$ 
for three years of increased $^{14}$C production.

Since the solar activity was very strong just before the $^{14}$C spike (both around AD 770 and 1790),
we could assume a $^{14}$C background production rate of only $\sim 1.2$ $^{14}$C atoms cm$^{-2}$ s$^{-1}$ 
(as during the 1990 solar maximum).
However, given the dearth of aurora reports for the time AD 774 to 785 (Neuh\"auser \& Neuh\"auser 2015a),
even though of many reports about other celestial events in that period and reports of aurorae
before and after that time, the solar activity was very low since the $^{14}$C spike (AD 774 to 785). 
Hence, we have to assume the solar minimum value of 
$\sim 2.2$ atoms cm$^{-2}$ s$^{-1}$ as during the very deep 2010 solar minimum (Kovaltsov et al. 2012).
Adding up the numbers (spike around AD 775-777 plus background value at solar minimum),
we obtain a $^{14}$C production rate of some $\sim 3.5 \pm 0.3$ $^{14}$C atoms cm$^{-2}$ s$^{-1}$ 
for three year {\em event} duration 
-- or $6 \pm 1$ $^{14}$C atoms cm$^{-2}$ s$^{-1}$ for only one year as {\em event duration}.

This value has to be compared to the largest possible $^{14}$C production rate,
which can be $\sim 3$ $^{14}$C atoms cm$^{-2}$ s$^{-1}$ --
or even $\sim 9$ $^{14}$C atoms cm$^{-2}$ s$^{-1}$, if the geomagnetic field would be switched off,
as estimated by Kovaltsov et al. (2012) 
-- some $50\%$ higher in other calculations (e.g. Lingenfelter 1963, Lal 1988).
For a geomagnetic field strength of $\sim 8.8 \cdot 10^{22}$ Am$^{2}$
around both AD 770s and 1790s (e.g. Korte et al. 2009, Korte \& Constable 2011), we can 
estimate the $^{14}$C production rate to be $\sim 0.8$ to $2.9$ atoms cm$^{-2}$ s$^{-1}$ 
for a solar modulation potential from 0 to 2000 MV (from Kovaltsov et al. 2012 figure 2),
again some $50\%$ higher in other calculations. 

Furthermore, one has to consider latitudinal effects: while the above numbers for
expected $^{14}$C production are all global averages, the production rate 
for geomagnetic latitudes above $\sim 40^{\circ}$ is higher 
than the global average (e.g. Masarik \& Beer 1999),
e.g. a factor of $\sim 2$ higher for $\sim 60^{\circ}$ geomagnetic latitude; 
we list here the geomagnetic latitudes for the trees measured for the AD 775 variation 
considering five reconstructions for the location of the geomagnetic 
pole (Arch3k by Bonadini et al. 2009, Cals3k3 by Korte et al. 2009, Cals3k4 by Korte \& Constable 2011, 
Cals10k1b by Korte et al. 2011, and Pfm9k1 by Nilsson et al. 2014):
\begin{itemize}
\item M12, Yaku Island, Japan, $\sim 30-34^{\circ}$; 
\item Usoskin et al. (2013), tree Steinbach 91, Germany, $\sim 50-56^{\circ}$; 
\item  Jull et al. (2014), California, $\sim 39-42^{\circ}$ 
and Siberia $\sim 71-77^{\circ}$; 
\item Wacker et al. (2014), M\"ustair, Switzerland, $\sim 47-53^{\circ}$; 
\item B\"untgen et al. (2014), Austrian alps, $\sim 48-54^{\circ}$; and 
\item G\"uttler et al. (2015): new Zealand Kauri, $\sim 36-41^{\circ}$ (south). 
\end{itemize}
Four to six of the seven tree locations are at higher geomagnetic latitudes 
than $\sim 40^{\circ}$, 
where more $^{14}$C production has to be expected than on the global average.

Since the estimated range ($\sim 3.5 \pm 0.3$ $^{14}$C atoms cm$^{-2}$ s$^{-1}$ production rate AD 775-777) 
is still not significantly larger than the largest possible value of
$\sim 3$ (or 9) $^{14}$C atoms cm$^{-2}$ s$^{-1}$ (global average, but larger for higher latitudes),
the effect suggested by us is possible quantitatively --
even without assuming a strong short-term geomagnetic field drop around the same time.

There are high uncertainties in all numbers involved: 
\begin{enumerate}
\item The estimated $^{14}$C production around AD 775
differs by a factor of 4-6 between M12 and Usoskin et al. (2013) depending on the carbon model.
\item While Usoskin et al. (2013) assumed that $70~\%$ of the $^{14}$C is produced in the 
stratosphere and $30~\%$ in the troposphere, Pavlov et al. (2013, 2014) used
$65~\%$ and $35~\%$, respectively,
\item An altitude change in the border between those two regions
or sudden injection of stratospheric air into the troposphere can lead to strong modifications.
\item When estimating the presumable extra production due to a flare, 
Usoskin et al. (2013) assumed some background production rate and
also that this background stayed constant for some 30 yr (carbon cycle);
however, if the cause for the $^{14}$C increase would have been a flare, i.e. increased
solar activity, the background (cosmic ray induced) radioisotope production should have
decreased (the Forbush decrease)
-- then, the observed $^{14}$C curve would be an even
larger deviation from the background, so that the $^{14}$C production rate and the
presumable solar super-flare would need to be even larger (and, hence, less likely) than assumed so far.
\item The calculations of $^{14}$C production rates also
suffer from numerical assumptions, e.g. on the cutoff rigidity and partly unknown 
cosmic ray and solar particle spectra and cross-sections,
and are quite different from team to team.
\item Reconstructions of the geomagnetic field for the
AD 770s are also highly uncertain and strongly different from team to team 
(e.g. Korte et al. 2009 compared to Knudsen et al. 2008).
\item As given above, also the pure $^{14}$C measurements in p.m. have large error bars:
\begin{itemize}
\item Fig. 6 shows the known inter-hemispheric offset;
\item Fig. 9 shows also inter-continental offsets;
\item Fig. 9d \& e show that there are also inter-laboratory offsets, e.g. differences
between the Mannheim and Z\"urich 
labs, which both 
measured the same German Oak tree
(Usoskin et al. 2013).
\end{itemize}
\end{enumerate}

Hence, given all these uncertainties, it can be stated that the effect suggested by us here
(strong rapid drop in solar wind, but not neccessarily down to zero) is quantitatively possible.

\begin{figure*}
\begin{center}
{\includegraphics[angle=270,width=15cm]{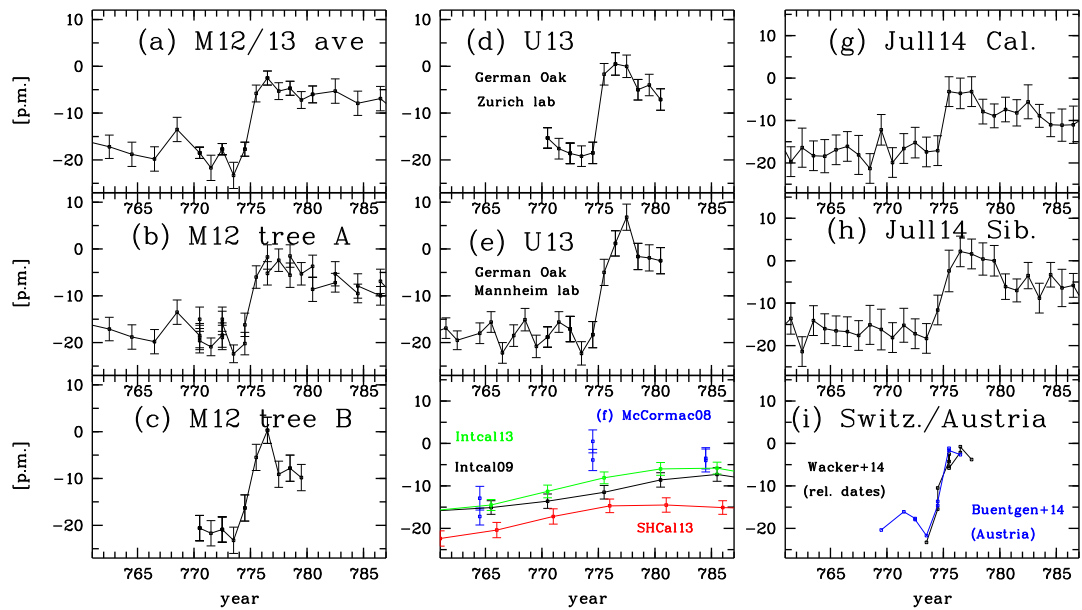}}
\caption{{\bf Variation of $^{14}$C around AD 775 as measured by different teams for different trees:}
(a) The average from M12 and Miyake et al. (2013a);
(b) just tree A from M12 and Miyake et al. (2013a);
(c) just tree B from M12;
(d) the German Oak tree from Usoskin et al. (2013) as measured in the Z\"urich lab;
(e) the same tree as in (d) as measured in the Mannheim lab;
(f) Irish Oak trees measured by McCormac et al. (2008) in blue
compared to Intcal09 data (Reimer et al. 2009) in black,
Intcal13 (Reimer et al. 2013) in green, and SHCal13 (Hogg et al. 2013b) in red;
(g) data from Jull et al. (2014) for a Californian tree;
(h) data from Jull et al. (2014) for a Siberian tree -- note that the 
three {\em yearly} increases from AD 773 to 776 are quite similar; and
(i) data from B\"untgen et al. (2014) in blue and Wacker et al. (2014) in black from 
Austria and Switzerland, respectively (the wood from the Swiss church does not have
absolute ages, Wacker et al. 2014); note that for the Wacker et al. (2014) data,
the increase from AD 773 to 774 is $+10.3$ p.m., while the increase from AD 774 to 775
is $+8.5$ p.m. (after averaging multiple data points for single years).
While the largest jump in $^{14}$C with 1- to 2-yr time resolution is often from AD 774 to 775,
the total increase from year to year lasts three to four years in the different data sets.
The rise in $^{14}$C in the 770s was first detected in McCormac et al. (2008)
in the two Irish Oak trees number Q9308 and Q9309.
The $^{14}$C data are plotted here exactly into their measurement year,
because here we do not compare them to other proxies.}
\end{center}
\end{figure*}

While the fact that the $^{14}$C increase around AD 775 lasted a few years may well
be consistent with an increased $^{14}$C production rate of only one year
or less (due to the carbon cycle, e.g. M12), such an increase over few years would
also be consistent with a decrease of solar activity and wind over a few years.
M12, Usoskin et al. (2013), and G\"uttler et al. (2015) came to the conclusion that the
$^{14}$C increase around AD 775 lasted only up to about one year by fitting their data
with a carbon cycle model -- with an event duraction of $\le 1$ yr being only slightly
more significant than, e.g., 2 or 3 yr (M12). Apart from the uncertainties in such models,
we would like to note that in the Usoskin et al. (2013) Z\"urich lab data and
in the G\"uttler et al. (2015) data, $^{14}$C increases by only 2 yr,
while in the other data sets (M12 trees A and B, Usoskin et al. (2013) Mannheim lab,
Jull et al. (2014) Siberian tree, Wacker et al. 2014, and B\"untgen et al. 2014), $^{14}$C increases by 3-4 yr,
so that a fit to those data would probably give a longer {\em event} duration.
Also note that in the Jull et al. (2014) Siberian tree and in Wacker et al. (2014),
the increase form AD 773 to 774 is (almost) as large as from AD 774 to 775.

Since the AD 775 and 1795 $^{14}$C variations are similar, one would also have to consider
the super-flare hypothesis for AD 1795, also at the border between two Schwabe cycles during
a solar activity minimum. Indeed, one could argue that the Sun released a large amount of
magnetic energy at the end of a Schwabe cycle (super-flare), so that the next cycle(s) are
weak (and possibly short), just as at the start of the Dalton minimum. 
However, neither super-strong aurorae nor extremly large spots were observed around AD 1795
nor around AD 775 (Neuh\"auser \& Neuh\"auser 2015a).
Such a behaviour was not observed at the starts of the other Grand Minima, neither at the start
of the Gleissberg minimum, nor at the starts of the Dark Age, Oort, Wolf, Sp\"orer,
or Maunder Minima.
No super-flares or very large spots or
aurorae were observed at the start of any Grand Minimum, also not in the telescopic time.

\section{The timing of large short-term variations}

We will now discuss whether the large short-term variations seen in $^{14}$C
around AD 775 and 1795 happen at a special moment of solar activity.
We will discuss the conclusions within the context of solar physics.

\subsection{1st and 2nd harmonics of the Schwabe cycle}

If the Gnevyshev-Ohl rule is due to the Hale-Babcock magnetic cycle,
and since the Gnevyshev-Ohl rule is fullfilled after the end of the Maunder
minimum until at least cycle no. 4 in the
AD 1790s,\footnote{It is fulfilled also afterwards, if cycle no. 4 is split into 4a and 4b}
the strong rapid drop in $^{14}$C in the 1790s happened at or around the turnover
from one Gnevyshev-Ohl pair and one Hale-Babcock magnetic pair to the next pair.

Then, there are also packages of (about) four Schwabe cycles with certain hemispheric
asymmetries: one solar hemisphere leads the spot formation and decays earlier
for most of the time for (about) four Schwabe cycles (e.g. Zolotova et al. 2009, 2010),
and also non-homogeneous hemispheric spot distributions,
which are due to a variation in the location of the magnetic equator
(e.g. Waldmeier 1957 for total spot numbers north and south,
and Pulkkinen et al. 1999 considering the heliographic latitudes of the spots).
The two effect are anti-correlated: when the northern side leads,
then there are more spots on the southern side, and viceversa --
at least it can be said that there is no evidence that this anti-correlation was disobeyed since
after the end of the Maunder Minimum until at least about 1965 (Temmer et al. 2006),
which can be said only for those times where sufficient data are known (Zolotova et al. 2009, 2010).

The hemispheric asymmetry effects are small and only slightly above the noise,
so that it is hard to date the turnover precisely and accurately. However, the data in
Waldmeier (1957), Pulkkinen et al. (1999), and Zolotova et al. (2009, 2010)
are fully consistent with the duration of those packages being four Schwabe
cycles and the turnovers being either during or after the 4th cycle.

Also, those packages of four cycles are in phase with the Gnevyshev-Ohl pairs and
the Hale-Babcock cycle, i.e. consist of two such pairs each.
Like the Hale-Babcock magnetic cycle is the 1st (sub-)harmonic of the Schwabe cycle,
i.e. always exists of two Schwabe cycles,
and like the packages of four Schwabe cycles (as 2nd (sub-)harmonic) 
consist of two Gnevyshev-Ohl and Hale-Babcock pairs each,
two packages of four Schwabe cycles with different hemispheric
asymmetries form a long cycle with 8 Schwabe cycles, which we call the {\em octave sequence},
the 3rd (sub-)harmonic of the Schwabe cycle.

\subsection{Octave sequence as 3rd harmonics}

We now have to consider the phase locking of the octave sequence.
This can be decided by considering whether, e.g., there is a stronger change in solar
activity before or after one of those two different kinds of packages of four Schwabe cycles.
Indeed, for the last three centuries, we can see a strong variation in solar activity
before those four Schwabe cycles,
where the northern side is leading and the southern side is stronger:
this is clearly seen in about AD 1712/15 (end of Maunder minimum before cycle -3),
about 1795 (before cycle 5 or 4b),
1879 (from 11 to 12), and 1965 (from 19 to 20).
At around 1795, 1879, and 1965, the sunspot and aurora numbers droped
strongly from one cycle to the next (also noticed by Richard 2004 for sunspot numbers);
at 1879 and 1965, also the aa-index decreased strongly.
At the end of the Maunder minimum, the number of spots and aurorae suddenly increased.
For all four cases, from cycle -4 to -3, from 4 (or 4a if split) to 5 (or 4b), from 11 to 12, 
and from 19 to 20, it can be stated that the activity proxies changed strongly.

\begin{figure}
\begin{center}
{\includegraphics[angle=270,width=8cm]{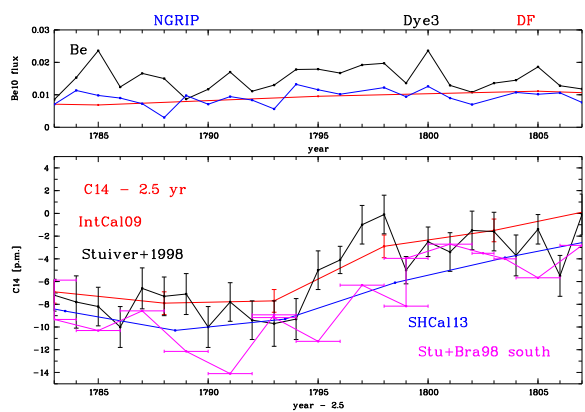}}
\caption{{\bf Increase of $^{14}$C and $^{10}$Be from AD 1783 to 1807:}
Top: $^{10}$Be in 0.01 atoms cm$^{-2}$ s$^{-1}$ from NGRIP (blue),
Dye 3 (black), and Dome Fuji (red) showing an increase from 1793 to 1794.
Bottom: $^{14}$C from Intcal09 (red) and Stuiver et al. (1998) in black with one year time-resolution
for the northern hemisphere and data from Stuiver \& Braziunas (1998) for the southern hemisphere in pink
(time range indicated by horizontal error bars) and SHCal13 in blue from Hogg et al. (2013b)
-- plotted 2.5 yr ahead of the measurement time as usual --
showing the strong fast increase from AD 1796 to 1800, but slightly earlier in the south.}
\end{center}
\end{figure}

The sinusoidal secular change in the location of the magnetic equator seen
in Pulkkinen et al. (1999) does not continue into the Maunder Minimum:
for the last four cycles of the Maunder Minimum, one cannot state that the magnetic
equator was shifted towards the north, the opposite is true (almost only spots in the south);
also, one cannot state that north was not leading.
Also for Richard's (2004), Waldmeier's (1957), and all kinds of
Gleissberg-like long cycles,\footnote{Gleissberg (1958) wrote that {\em one 80-year cycle, 
on the average, equals to 7.1} Schwabe {\em cycles ...
The actual length of the 80-year cycles semm to vary considerably} from $\sim 25$ to 125 yr (his figure 2).
In Gleissberg (1967), he noticed that the long cycles last for six to eight cycles (from 1632-1944).
The {\em Gleissberg cycles} according to Gleissberg himself would last for $\sim 80$ yr on average 
(with a large range of 25 to 125 yr), but not neccessarily made up of exactly eight cycles.} 
the Maunder Minimum appears to be an exception.
For our picture above, the turnover from the Maunder Minimum to afterwards
does not need to be an exception:
there is not neccessarily a large {\em drop} in activity before each of those four packages where
south is stronger, but a large {\em change} (drop or rise).
While in the last three full octave sequences (cycles -3 to 19), there was first
the package of four Schwabe cycles with south being stronger and then four cycles with
north being stronger, this may well have been the opposite in the Maunder Minimum --
possibly due to a polarity change in some magnetic field
(which may even be one cause for a Grand Minimum).

In those four Schwabe cycles where the magnetic equator in shifted towards the south,
i.e. when north is leading, the activity proxies are generally weaker;
this holds definitely during cycles 12-15, the Gleissberg minimum,
and probably also for cycles 5-7 (or 4b to 7), the Dalton minimum, see Zolotova et al. (2009).
In cycles 20-23, 
when the southern side had a similar strength than the northern side,
the total activity was still relatively strong;
in the last four Schwabe cycles of the Maunder minimum,
when south was much stronger than north (almost no spots in the north),
the total activity was even very weak.
As far as the Dalton and Gleissberg minima are concerned, the change in hemispheric
asymmetry (i.e. from one package of four cycles to the next, or even from one
octave sequence to the next) happened in about the middle or in the last half of the
previous 8th (last) Schwabe cycle, i.e. in cycle 4 (or 4a) before Dalton and in cycle 11 before Gleissberg;
here, the southern side suddenly started to be stronger than the north.
Thus, the turnover may not happen in the activity minimum, but around the
polarity change, the turnover from one Hale-Babcock pair to the next.

At the octave sequence turnover from cycle 19 to 20, there was an
{\em outstanding excess of the northern hemisphere} (Temmer et al. 2006
considering the sunspot numbers).
This turnover was different from the two others (before Dalton and Gleissberg)
in the sense that the northern side 
remained strong after 
the end of the 8th cycle
(for cycles 20-23, the total northern and southern sunspot numbers, added up over
those four cycles, was similar);
according to Pulkkinnen et al. (1999), the magnetic equator (determined from the
average heliographic latitudes of spots) moved from the 
northern to the southern hemisphere around the turnover from cycle 19 to 20.
Interestingly, opposite to Dalton and Gleissberg,
the new octave sequence (since cycle 20) started with four relatively strong cycles
(and may end with four weak cycles, the current Grand Minimum);
the prolonged deep minimum around 2008/2009 may be connected
to the turnover from one package of four cycles to the next.

If the Gnevyshev-Ohl rule is due to the Hale-Babcock cycle of a certain (main) magnetic field,
then the packages of four Schwabe cycles with a certain hemispheric asymmetry
may well be due to another magnetic cycle.
If both these cycles have their turnover at the same time, the effects add up.
Due to some symmetry or feedback mechanism, this effect may be particulary strong after
activity increased over eight Schwabe cycles to a very high level, as seen before the AD 770s and 1790s.
At those times, solar activity droped strongly and suddenly.
We can also see a strong change in solar activity
proxies after the 8th cycle in 1879 and 1965 (e.g. Richard 2004).
The effect is not always as strong as it was in the AD 1790s and 770s;
it may well be particulary strong, when solar activity was particulary strong
in the last few of the previous eight Schwabe cycles (and/or after $\le 200$ yr of
stronger-than-normal solar activity variations).
We would then expect that the turnover from one octave sequence (ending with
strong activity) to the next sequence (starting with weak activity) took place
not only around AD 1795 (with the strong radiocarbon rise), but also around AD 775 and BC 671.

The additional relatively strong radiocarbon increase at AD 993/4 (Miyake et al. 2013a)
may also have been at the turnover from one such package of four Schwabe cycles to the next:
from AD 774/5 to 993/4, we have roughly 220 yr, i.e. three (short) long cycles or
octave sequences; the fact that those three long cycles were relatively short may well
be due to the fact that we were in the Medieval Warm period of increased solar activity.

\section{Summary}

We compared the $^{14}$C variation around AD 775 with two similar cases around AD 1795 and BC 671.
We could show that all these cases show a similar pattern of secular variation in $^{14}$C:
\begin{enumerate}
\item First a Grand Maximum for some 80 yr (Dark Age and Maunder Minimum, respectively),
\item then eight Schwabe cycles with increasingly strong solar activity with very low $^{14}$C at the end
(due to very strong solar activity),
\item then a sudden strong rise in $^{14}$C at the end of the last of those eight Schwabe cycles,
i.e. a sudden transition from very strong to very low activity,
\item a rise of $^{14}$C up to a high value, because activity stayed low afterwards for a few Schwabe
cycles (e.g. the Dalton Minimum) -- after those weak cycles, activity recovers by increasing significantly.
\end{enumerate}
The secular evolution lasts in total up to some 200 yr,
the strongest rises and drops are within some 60 yr.

The three cases studied are the only ones within the period from BC 1000 to AD 1900,
so that one could consider the rate of such variations to be about once per millenium.

Around AD 1795, we can state that the sudden drop in activity (the sudden start of the
Dalton minimum) happened after a Hale-Babcock and a Gnevyshev-Ohl pair -- and also after
a package of four Schwabe cycles with the same hemispheric asymmetry
and after two such packages of four Schwabe cycles after the end of the Maunder Minimum:
at the turnover from one octave sequence to the next one, strong activity changes can happen. 
This was probably also the case for the $^{14}$C variation around AD 775,
which happened eight Schwabe cycles after the end of the Dark Age Grand Minimum.

The strong increase in $^{14}$C around AD 775 and AD 1795 lasted a few years.
Those strong increases happen when we are at the same time
\begin{enumerate}
\item in the declining phase of a strong Schwabe cycle, 
\item at the end of a longer period with quite high activity (Grand Maximum), and
\item at the start of a period with low activity (e.g. Dalton minimum).
\end{enumerate}

We do not need to explain the $^{14}$C variation around AD 775 with a solar super-flare:
\begin{itemize}
\item the flare would be as large as never observed otherwise on the Sun,
\item as large as never confirmed on a solar analog star (Neuh\"auser \& Hambaryan 2014), and
\item there is no evidence for strong aurorae nor large sunspots in the 770s (Neuh\"auser \& Neuh\"auser 2015a).
\end{itemize}
Also, there is no evidence for
a rare extra-solar (Galactic) gamma-ray burst.\footnote{Pavlov et al. (2013, 2014) expect a
short recombination airglow in 30-40 km height with 4-5 $\times 10^{6}$ erg cm$^{-2}$ fluence
after a gamma-ray burst, but there is no such evidence:
the observations in the Chronicle of Zuqnin for AD 772 and 773 (and maybe also 774) clearly
refer to normal aurora curtains with red and yellow color, 
and a report from the Ulster Chronicle for Ireland for AD 772 Sep 29 ({\em The hand-clapping on St Michael's 
Day which called fire from heaven.}) also refers to an aurora ({\em fire from heaven}), if a celestial event,
see Neuh\"auser \& Neuh\"auser (2015a) for discussion (aurorae are usually above 100 km);
a gamma-ray burst would also have led to strong ozone layer depletion and wheather disturbences
(e.g. Pavlov et al. 2014), which are not reported (at least for the northern hemisphere).} 
We can instead explain the $^{14}$C variation as more 
or less normal solar activity 
variation (three strong cases observed in 2900 yr),
namely with a modulation of (extra-solar) cosmic rays by solar activity variation.

{\bf Note added in proof.}

Most recently, Miyake et al. (2015) and Sigl et al. (2015) obtained quasi-annual $^{10}$Be data
around AD 775, both showing strong fluctuations.
Miyake et al. (2015) obtained data for Antarctica and considered whether the largest peak
in AD 779/780 ($\sim 2 \times$ above the strongly fluctuating background
noise in the other years, 763-794) could be the $^{10}$Be signal
corresponding to the $^{14}$C signal around AD 775, but it would be 4 yr too late.

Sigl et al. (2015) took data for three Arctic sets (NGRIP, NEEM, and TUNU)
and one Antarctic set (WDC), all showing the strongest peak in AD 768, 
presumably 7 yr too early.
They do not compare their data with Miyake et al. (2015).
The four data sets presented in Sigl. et al. (2015) show strong differences
in most years, when comparing one set with another, e.g. in their figure 1 b,
local NEEM minima around 776, 780, and 787 correspond to local NGRIP maxima.
The yearly mean\footnote{without the data point for 769.5 in TUNU,
which we did not want to assign arbitrarily to either the 769 or the 770 year --
if we would include this data point with low $^{10}$Be flux in taking the means,
the peak in Fig. 11 would be even less significant} 
of the Sigl et al. (2015) set-to-set deviations
(measured per year) is $3.0 \pm 2.1 \times 10^{3}$ atoms/g, 
which we interprete as additional systematic error. 
The mean of the error bars given in Sigl et al. (2015)
are 0.52, 1.37, 0.79, and $0.68 \times 10^{3}$ atoms/g for NEEM, NGRIP, TUNU, and WDC,
respectively, i.e. some four times smaller.

We plot the yearly means (obtained from the four sets from Sigl. et al. 2015)
with the yearly fluctuation as error bars in Fig. 11, where we can see
that the strongest peak in AD 768 ($38.5 \pm 5.1 \times 10^{3}$ atoms/g)
is only $2 \sigma$ above the background noise of the time before and after the peak, 
the mean for the time outside of AD 766 to 772 is $16.9 \pm 5.4 \times 10^{3}$ atoms/g.

\begin{figure}
\begin{center}
{\includegraphics[angle=270,width=8cm]{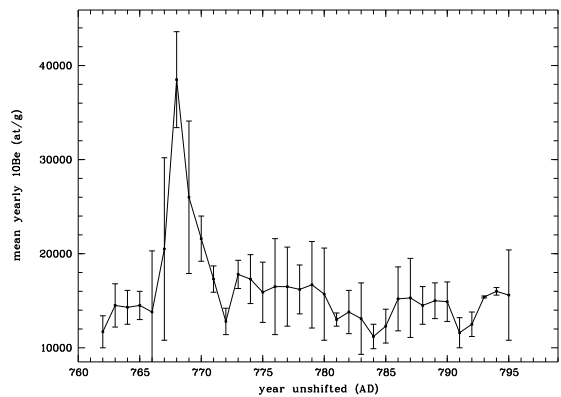}}
\caption{{\bf Mean yearly $^{10}$Be data from Sigl et al. (2015).}
We plot for each year the mean from the four data sets in Sigl et al. (2015)
with the error of that mean as systematic error bars. They are $\sim 4$ times 
larger than the measurement errors given in Sigl et al. (2015).
The largest peak, seen here in AD 768, is now only $\sim 2\sigma$ above
the remaining background noise. This is similar in Miyake et al. (2015) $^{10}$Be
data, but with the largest peak being in AD 779.  }
\end{center}
\end{figure}

The $^{10}$Be signal in Sigl et al. (2015) is 7 yr too early, and it is only $2 \sigma$ above the noise. 
The TUNU data set does not even cover AD 774/5, 
so that it is not clear whether the peak
seen in AD 768 is the largest one (data only from 763-770).
For the NEEM data set around AD 993/4, they took data for only 13 years,
so that the identification of the peak presumably related to the $^{14}$C variation
around AD 993/4 is also uncertain (with the same systematic error bars as around AD 775,
it would be much less significant). 

In the yearly $^{10}$Be data of the last four centuries,
there are more deviations of similar strength 
(as claimed by Sigl et al. (2015) for AD 775 and 993/4), 
e.g. around 1460, 1605, 1865, and 1890 in NGRIP and/or Dye-3,
all unrelated to strong $^{14}$C variations.
As long as the cause for the $^{14}$C variation around AD 775 and 993/4 is
not certain, it is uncertain whether a $^{10}$Be increase should be expected at all.

Even if a $^{10}$Be signal would be expected, this may not neccessarily be in AD 775,
because the $^{14}$C increase around AD 775 took 3-4 yr 
in most data sets\footnote{Also 
note that in the Jull et al. (2014) Siberian tree and in Wacker et al. (2014),
the increase form AD 773 to 774 is (almost) as large as from AD 774 to 775.}.
Furthermore, the atmospheric residence time and the incorporation of $^{14}$C into 
tree rings is $\sim 20$-25 yr (with the largest yearly $^{14}$C incorporation rate
in the third year, Sigman \& Boyle 2000),
while $^{10}$Be is incorporated mostly within the first year (Heikkil\"a et al. 2008);
hence, one might expect a time difference between such peaks.

We can now estimate the $^{14}$C to $^{10}$Be production ratio around AD 774/5:
for $^{14}$C, there was a production of S$_{C14} = 19 \pm 4$ atoms cm$^{-2}$ s$^{-1}$ (M12,
for one year), but 4-6 times less according to Usoskin et al. (2013) --
a signal above the $^{14}$C background rate of B$_{C14} \simeq 2$ atoms cm$^{-2}$ s$^{-1}$.
The above given peak in $^{10}$Be in the yearly means (S$_{Be10} = 38.5 \pm 5.1 \times 10^{3}$ atoms/g
in one year, AD 768) is 2.28 times larger than the background noise outside the peak in Sigl et al. (2015) data;
for a $^{10}$Be background level of B$_{Be10} \simeq 0.031$ atoms cm$^{-2}$ s$^{-1}$ (Kovaltsov \& Usoskin 2010), 
one can estimate the $^{10}$Be signal in atoms cm$^{-2}$ s$^{-1}$.
We assume that the $^{10}$Be peak in AD 768 corresponds to the $^{14}$C peak around AD 775
(here using 4-6 times less $^{14}$C than in M12 as recommended by Usoskin et al. 2013). 
Then, the ratio is

\begin{displaymath}
\frac{S_{C14} - B_{C14}}{S_{Be10} - B_{Be10}}~~~=~~~29~~~\mbox{to}~~~69~~~\mbox{(mean 45)}
\end{displaymath}

This ratio is not consistent with the expectation for a solar flare,
where only 15-17 times more $^{14}$C is expected than $^{10}$Be (Pavlov et al. 2013, 2014).
The value obtained from the Miyake et al. (2015) $^{10}$Be data is
within the above range: the peak at AD 779 is
$\sim 2$ times above the background noise outside that peak.

When shifting the strongest Miyake et al. (2015) and Sigl. et al. (2015) ($2 \sigma$) peak in $^{10}$Be
to the same year, and when considering the systematic errors in Sigl et al. calculated above,
then those data sets are not inconsistent with each other: The amplitude of the strongest peak in
Miyake et al. (2015) is only two times larger than the year-to-year (background) fluctuations
(which are also much larger than the formal error bars given).

The $^{10}$Be data from both Miyake et al. (2015) and Sigl. et al. (2015)
are not inconsistent with a strong rapid drop in solar
activity around AD 775: 
not only $^{14}$C should have increased within a few years
from before AD 775 to after AD 775, but also $^{10}$Be.
In all quasi-annual data sets, the $^{10}$Be flux after the peak (which those
authors contribute to the presumable AD 775 $^{14}$C peak by shifting the
$^{10}$Be data in time) is similar or larger than before that peak --
unfortunately, the systematic error bars are too large and the time span 
covered before the peak are too small to show a statistical significant difference.
Around AD 1795, $^{10}$Be was significantly smaller before 1795 
than after 1795, as shown in Sect. 4.1.3 (similar also in figure 15
in Usoskin et al. 2015 with NGRIP and Dye-3).

While the tree-ring anomalies (N-Tree) and the global
temperature reconstruction (PAGES-2k) shown in Sigl. et al. (2015)
do not reflect all Grand Minima and Maxima, they do show both 
a strong temperature increase for several decades before AD 775
and a strong depression after about AD 775 for a few decades,
which would not be expected for a solar flare, but it is similar 
in both regards before and during the Dalton minimum 
(which started well before the 1815 vulcano eruption) -- 
maybe both temperature depressions are due to strong rapid drops in solar activity
(both after several decades of increaing activity). 

\acknowledgements
We obtained solar data and radioisotope data from www.ngdc.noaa.gov, www.radiocarbon.org, and ftp.ncdc.noaa.gov.
We would like to thank J. Beer for the $^{10}$Be GRIP and Dye 3 data in electronic form.
We retrieved the NGRIP $^{10}$Be from the IGBP PAGES/World Data Center for Paleoclimatology.
$^{14}$C data from Miyake et al. (2013ab) were received from F. Miyake in electronic form.
We could get the $^{14}$C data (in p.m.) from McCormac et al. (2008) from P. Reimer.
We thank P. Reimer, A. Hogg, and J. Palmer also for further advise on the radiocarbon data.
$^{14}$C data from Usoskin et al. (2013) and Wacker et al. (2014) were received in electronic form from L. Wacker.
We thank J.M. Vaquero for the historic naked-eye sunspot record (Vaquero et al. 2002) in electronic form.
We thank M. Korte (GFZ Potsdam) for advise on the geomagnetic field.
We thank T. Krejcova (now U Hamburg) for providing us with the paper by Krivsky \& Pejml.
We used the Silverman aurora catalog on nssdcftp.gsfc.nasa.gov/miscellaneous/aurora.
We acknowledge R. Arlt for a lot of useful discussion.
T. Zehe and F. Gie\ss ler (U Jena) provided magnetic field calculations.
We also thank V. Bothmer for discussion about strong solar storms.
We are grateful to an anonymous referee for advise which also improved the 
structure and readabilty of the paper.

\end{document}